\documentclass[twocolumn,showpacs,superscriptaddress,floatfix,amsmath,amssymb]{revtex4}

\usepackage{bm}
\usepackage{graphicx}
\usepackage{dcolumn}
\usepackage{bm}
\usepackage{braket}
\usepackage{array}
\usepackage{multirow}
\usepackage{url}
\usepackage{tabularx}
\usepackage[bookmarks=true,colorlinks=true,urlcolor=blue,linkcolor=blue,citecolor=blue]{hyperref}
\usepackage{longtable}
\usepackage{makecell}
\usepackage{siunitx}
\usepackage[dvipsnames]{xcolor}
\usepackage[normalem]{ulem}
\begin{document}

\preprint{APS/123-QED}

\title{Competing excitation quenching and charge exchange \\  in ultracold Li-Ba$^+$ collisions}

\author{Xiaodong Xing}
\affiliation{Universit$\acute{\text e}$ Paris-Saclay, CNRS, Laboratoire Aim$\acute{\text e}$ Cotton, Orsay, 91400, France}
\affiliation{Department of Physics, University of Nevada, Reno, NV 89557, USA}
\author{Pascal Weckesser}
\affiliation{Albert-Ludwigs-Universit$\ddot{a}$t Freiburg, Physikalisches Institut, 79104 Freiburg, Germany}\affiliation{Max-Planck-Institut für Quantenoptik, 85748 Garching, Germany}\affiliation{Munich Center for Quantum Science and Technology (MCQST), 80799 Munich, Germany}
\author{Fabian Thielemann}
\affiliation{Albert-Ludwigs-Universit$\ddot{a}$t Freiburg, Physikalisches Institut, 79104 Freiburg, Germany}
\author{Tibor Jónás}
\affiliation{University of Debrecen, Doctoral School of Physics, Egyetem tér 1., 4032 Debrecen Hungary}
\affiliation{HUN-REN Institute for Nuclear Research (ATOMKI), Bem tér 18/c, 4026 Debrecen, Hungary}
\author{Romain Vexiau}
\affiliation{Universit$\acute{\text e}$ Paris-Saclay, CNRS, Laboratoire Aim$\acute{\text e}$ Cotton, Orsay, 91400, France}
\author{Nadia Bouloufa-Maafa}
\affiliation{Universit$\acute{\text e}$ Paris-Saclay, CNRS, Laboratoire Aim$\acute{\text e}$ Cotton, Orsay, 91400, France}
\author{Eliane Luc-Koenig}
\affiliation{Universit$\acute{\text e}$ Paris-Saclay, CNRS, Laboratoire Aim$\acute{\text e}$ Cotton, Orsay, 91400, France}
\author{Kirk W. Madison}
\affiliation{Department of Physics and Astronomy, University of British Columbia, Vancouver, BC V6T1Z1, Canada}
\author{Andrea Orbán}
\affiliation{HUN-REN Institute for Nuclear Research (ATOMKI), Bem tér 18/c, 4026 Debrecen, Hungary}
\author{Ting Xie}
\affiliation{State Key Laboratory of Molecular Reaction Dynamics, Dalian Institute of Chemical Physics, Chinese Academy of Sciences, Dalian, Liaoning 116023, People’s Republic of China}
\author{Tobias Schaetz}
\affiliation{Albert-Ludwigs-Universit$\ddot{a}$t Freiburg, Physikalisches Institut, 79104 Freiburg, Germany}
\affiliation{EUCOR Centre for Quantum Science and Quantum Computing, Albert-Ludwigs-Universität Freiburg, 79104, Germany}
\author{Olivier Dulieu}
\affiliation{Universit$\acute{\text e}$ Paris-Saclay, CNRS, Laboratoire Aim$\acute{\text e}$ Cotton, Orsay, 91400, France}

\date{\today}

\newcommand{\dth}{$5D_{3/2}$}
\newcommand{\dfh}{$5D_{5/2}$}
\newcommand{\soh}{$6S_{1/2}$}
\newcommand{\li}{$^6$Li}
\newcommand{\ba}{$^{138}$Ba$^+$}

\begin{abstract} 
Hybrid atom-ion systems are a rich and powerful platform for studying chemical reactions, as they feature both excellent control over the electronic state preparation and readout as well as a versatile tunability over the scattering energy, ranging from the few-partial wave regime to the quantum regime. In this work, we make use of these excellent control knobs, and present a joint experimental and theoretical study of the collisions of a single $^{138}$Ba$^+$ ion prepared in the $5d\,^2D_{3/2,5/2}$ metastable states with a ground state $^6$Li gas near quantum degeneracy. We show that in contrast to previously reported atom-ion mixtures, several non-radiative processes, including charge exchange, excitation exchange and quenching, compete with each other due to the inherent complexity of the ion-atom molecular structure.  We present a full quantum model based on high-level electronic structure calculations involving spin-orbit couplings. Results are in excellent agreement with observations, highlighting the strong coupling between the internal angular momenta and the mechanical rotation of the colliding pair, which is relevant in any other hybrid system composed of an alkali-metal atom and an alkaline-earth ion.

\end{abstract}

\pacs{....}
\maketitle

\section{Introduction}

Quantum mixtures of ultracold atomic gases are powerful platforms for opening new perspectives on dilute and condensed matter physics, due to their exquisite level of control that can be achieved in the experiments. Using neutral particles, various arrangements can be studied as for instance mixtures of quantum gases of different species, of identical species but in different quantum states, or immersion of a single impurity inside a quantum gases \cite{baroni2024}. Another promising mixed system recently emerges in the form of hybrid traps, \textit{i.e.} the merger of a single or a few laser-cooled and trapped atomic ions and an ultracold quantum atomic gas \cite{lous2022,deiss2024}. Such platforms offer opportunities to investigate quantum effects in ultracold ion-atom interaction, ultracold chemistry, formation of ultracold molecular ions for precision measurements, dynamics of a charged impurity in a neutral gas. An immediate question is raised: how stable are these hybrid systems? Due to the range of the ion-neutral interaction (varying as $R^{-4}$, with $R$ the interparticle distance) being much longer than the neutral-neutral interaction (varying as $R^{-6}$), three-body recombination events involving an ion and two neutrals is likely to occur \cite{krukow2016a,dieterle2020,mohammadi2021,perez-rios2021}, hampering the stability of such hybrid systems. However considering the inherent many-body nature of these mixtures, other perspectives can be envisioned, like the solvation of an ion within the atomic bath \cite{chowdhury2024}, or the formation of ion-atom complexes assisted by the trap potential \cite{hirzler2023}. 

The ion-neutral physics intrinsically depends on the details of the two-body interactions. Focusing on ion-atom hybrid systems, a wealth of experiments have been developed with various combinations, either homonuclear ones \cite{grier2009,haerter2012,dutta2018}, or heteronuclear pairs of alkali-metal (AM) atoms an alkaline-earth (AE) ions (or Yb$^+$) \cite{smith2005,schmid2010,hall2011,ratschbacher2012,haze2013,hall2013b,joger2017,sikorsky2018,mills2019,li2020,schmidt2020}. As the laser-cooling scheme involves the metastable state of the alkaline-earth ion, the hybrid trap offers access to the collisional dynamics of ion-atom systems in the electronically excited states, with high internal energy disposal, opening the possibility for charge exchange (CE) \cite{hall2013a,saito2017,joger2017}. Numerous recent experiments revealed that excitation exchange between the two particles is the dominant channel \cite{benshlomi2020}. Simplified collisional calculations using high-level electronic structure of the related molecular complex [AM-AE]+ have been used to elucidate this diversity for various systems like LiCa$^+$ \cite{saito2017}, RbSr$^+$ \cite{benshlomi2020}, RbBa$^+$ \cite{mohammadi2021}, RbCa$^+$ \cite{xing2022}. But observed scattering rates are still missing a full quantitative interpretation, emphasizing that more elaborated dynamical models must be employed.

In this work, we focus on an ion-atom combination, a $^{138}$Ba$^+$ ion interacting with $^6$Li atoms, with a large mass imbalance suitable for reaching the quantum regime of ultracold collisions \cite{cetina2012,feldker2020}. Among all such pairs of alkaline-earth ions and alkali-metal atoms, the entrance ground-state scattering channel Ba$^+$+Li has the lowest energy, such that the system is protected against radiative charge exchange (the channel Ba+Li$^+$ is closed). This feature is particularly suitable for the observation of magnetic Feshbach resonances (MFRs) \cite{tomza2015,weckesser2021b,thielemann2024}, a crucial step toward the quantum control of the collision, the formation of ultracold molecular ions \cite{hirzler2022} and for ultracold chemistry \cite{lous2022}. Adding internal energy in the particles by electronic excitation enriches the multiplicity of dynamical pathways to be investigated. In particular, the laser cooling scheme of the ion involves the lowest metastable state, namely $5d\,^2D$ for the Ba$^+$ ion, allowing for the observation of excited ion collisions with neutral atoms. Such collisions have been observed in various systems, and were initially thought to be dominated by charge exchange \cite{hall2011,hall2013a,hall2013b,haze2015,saito2017,joger2017,li2020}, but these collisions also exhibit an interplay between other strong scattering channels leading to excitation exchange and quenching \cite{benshlomi2020,mohammadi2021}. 

\begin{figure}[]
 \centering
 \includegraphics[width=8.0 cm]{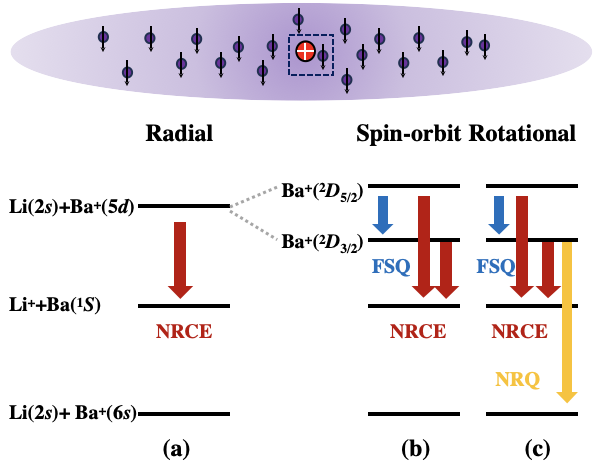}
 \caption{Schematic diagram of the energy levels of the [Li,Ba]$^+$ pair relevant for the present work, revealing the possible processes (non-radiative charge exchange (NRCE), non-radiative quenching (NRQ), fine-structure quenching (FSQ)) as the hierarchy of internal couplings are considered: dynamical radial coupling for (a), (b) and (c), the spin-orbit coupling for (b) and (c), and the rotational coupling for (c).}
 \label{fig:Elevels}
\end{figure}

In our system, we observe yet a different dynamical pattern. Several processes are found to compete with each other with comparable rates, as illustrated in Fig. \ref{fig:Elevels}: the non-radiative charge exchange (NRCE) 
\begin{eqnarray}
\text{Li}(2s\,^2S_{1/2}) &+& \text{Ba}^+(5d\,^2D_{3/2,5/2}) \to \nonumber \\
&\to&\text{Li}^+ + \text{Ba}(6s^2\,^1S),
\label{eq:NRCE}
\end{eqnarray}
the non-radiative quenching (NRQ) 
\begin{eqnarray}
\text{Li}(2s\,^2S_{1/2}) &+& \text{Ba}^+(5d\,^2D_{3/2,5/2})  \to \nonumber \\
&\to&\text{Li}(2s\,^2S_{1/2}) + \text{Ba}^+(6s\,^2S_{1/2}),
\label{eq:NRQ}
\end{eqnarray}
and the fine-structure quenching (FSQ)
\begin{eqnarray}
\text{Li}(2s\,^2S_{1/2}) &+& \text{Ba}^+(5d\,^2D_{5/2}) \to \nonumber \\
&\to&\text{Li}(2s\,^2S_{1/2}) + \text{Ba}^+(5d\,^2D_{3/2}).
\label{eq:FSQ}
\end{eqnarray}
In the rest of the paper we adopt the shortened notations Li($2S_{1/2}$), Ba$^+$($5D_{3/2}$), Ba$^+$($5D_{5/2}$), and Ba($^1S$) for the atomic states. Occasionally, the entrance channel in Eqs. \ref{eq:NRCE}-\ref{eq:NRQ} will be referred to as S+D, the outgoing channel in Eq. \ref{eq:NRCE} as Ion+S, and the outgoing channel in Eq. \ref{eq:NRQ} as S+S.

The paper is structured as follows. We first recall in Section \ref{sec:exp} the main features of the experimental setup and the observed rates for the various processes. In Section \ref{sec:structure-LZ} we present the electronic  structure of the LiBa$^+$ molecular ion, including our computed potential energy curves (PECs) and spin-orbit couplings (SOCs), and characterize their main features in terms of a simple Landau-Zener dynamical model, which is found insufficient to interpret the observations. Thus in Section \ref{sec:FCQS-MCQS} we propose two quantum scattering models including SOCs with and without the rotational (Coriolis) coupling, confirming the interplay between them, as it was anticipated in \cite{benshlomi2020}. Additional information regarding experimental setup and theoretical methods are provided in the Appendix.

\section{Experimental results}
\label{sec:exp}

In the hybrid setup in Freiburg, we combine a segmented linear Paul trap with an all-in-one-spot ultracold atom apparatus. A detailed description of the setup and various techniques has been presented in previous work~\cite{weckesser2021a,weckesser2021b} and are further elaborated in Appendix \ref{app:exp-protocol}, \ref{app:langevin}, \ref{app:d52prep}.

\begin{figure}[]
 \centering
 \includegraphics[width=0.6\columnwidth]{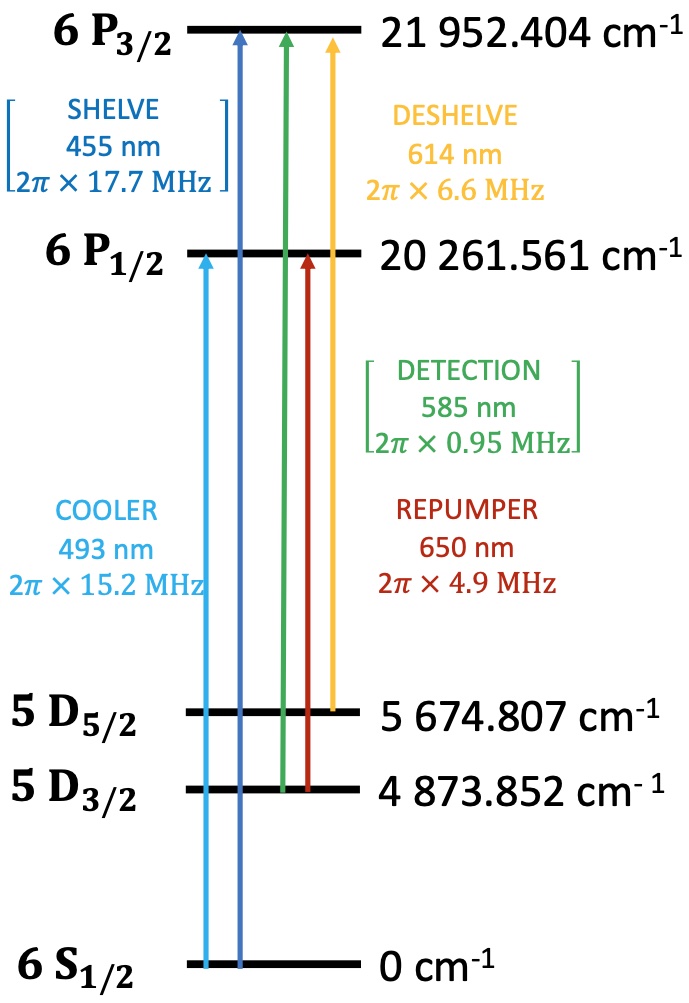}
 \caption{Energy levels of $^{138}$Ba$^+$, and relevant laser wavelengths for its cooling and its detection (see Appendix \ref{app:exp-protocol}). The brackets refers to lasers which were not available at the time of the present experiment.}
 \label{Ba+levels}
\end{figure}

At the beginning of each experimental sequence, we deterministically capture and prepare individual \ba{} ions~\cite{leschhorn2012,weckesser2021b}, by cooling them close to the Doppler temperature $T_{\mathrm{D}}\sim \SI{365}{\micro\kelvin}$. We compensate for radial and axial stray electric fields down to $\lesssim \SI{5}{\milli\volt\per\meter}$. For the interaction with the \li{} atoms we then either prepare the ion in the \dth{} or \dfh{} electronic manifold (Fig~\ref{fig:Elevels} and Fig~\ref{Ba+levels}). Their respective radiative lifetime is \SI{80}{\second} and \SI{32}{\second}, which are orders of magnitude longer than the duration of the experimental sequence, so that any change of internal state is induced by collisions. Once prepared, we then shuttle the ion along the axial direction to subsequently prepare the Li cloud.

For the \li{} atoms, we first load a magneto-optical trap (MOT) at the center of the Paul trap. We then transfer the atoms into a crossed optical-dipole trap (xODT) and perform evaporative cooling at higher magnetic fields close to quantum degeneracy. After evaporation we prepare the Li atoms in the level that correlates to the $\ket{f = 1/2, m_f = -1/2}$ sublevel at zero magnetic field, where $f$ is the Li total angular momentum (including electronic and nuclear spin) and $m_f$ its projection onto the quantization axis. Maintaining the magnetic field $B = \SI{293}{G}$ we then transfer the ion back to the trap center where the atoms reside. After an interaction duration $t_int$, we probe the resulting ion state, distinguishing between direct ion detection, a hot ion, an ion in the $5D_{5/2}$ state or loss of the ion from the trap (Appendix \ref{app:langevin}).

\begin{figure}[t]
 \centering
 \includegraphics[width=8.0cm]{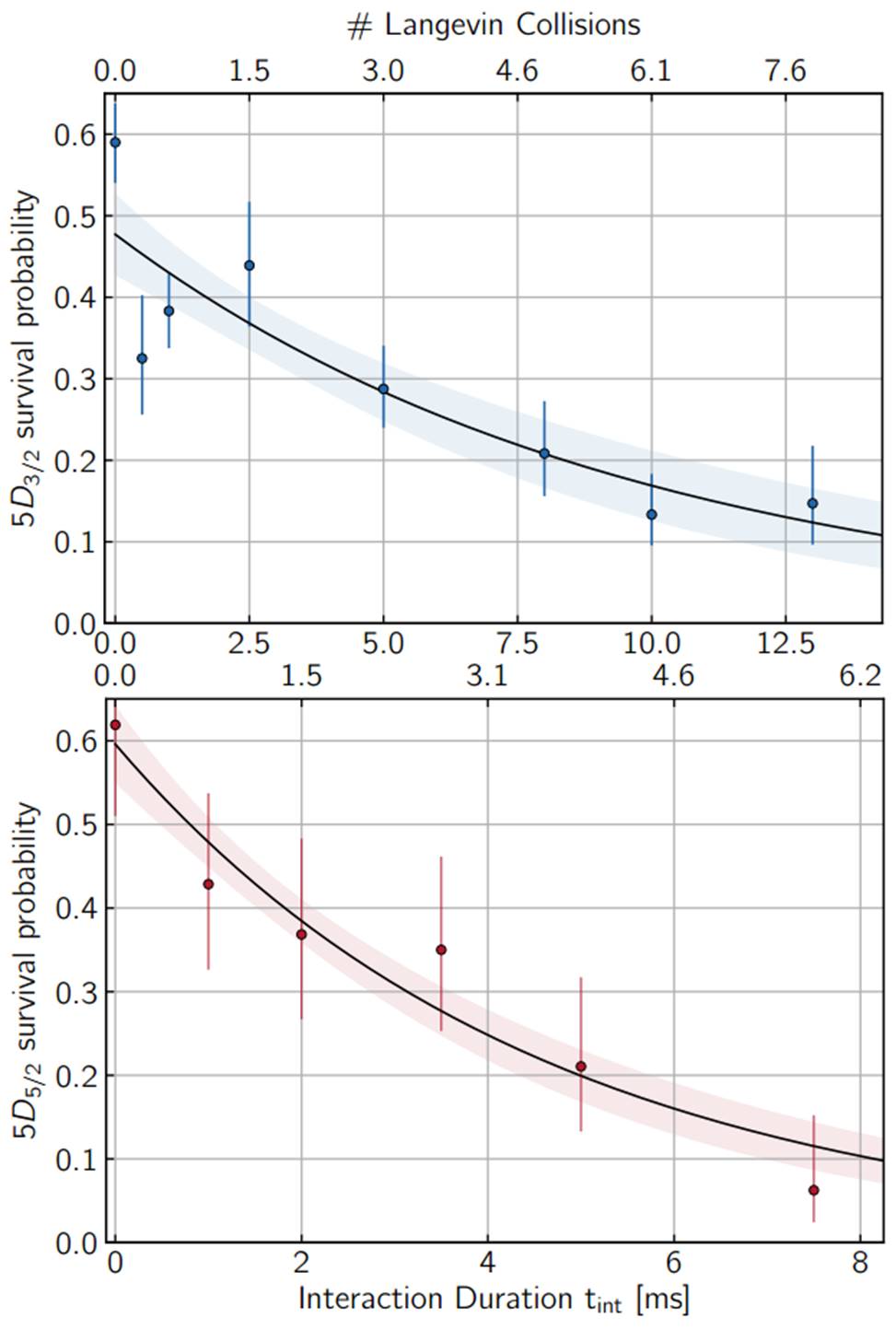}
 \caption{Survival probability of single Ba$^+(5^2D_{3/2})$ (top panel) or Ba$^+(5^2D_{5/2})$ (bottom panel) ion as a function of the interaction duration with the Li atoms, expressed in ms and in terms of number of Langevin collisions (\ref{app:langevin}).  Each point corresponds to an average over at least 20 events, while the error bars represent the 1$\sigma$-confidence interval. The black curve is an exponential fit, and the shaded area indicates the uncertainty of the fit. The non-unity survival probability for short duration is due to ion losses that occur during its movement through and interaction with the finite-size atomic cloud on its way to the center of the trap.}
 \label{fig:losses}
\end{figure}

Typical experimental measurements are displayed in Fig. \ref{fig:losses} for both $5D_{3/2}$ and $5D_{5/2}$ Ba$^+$ state preparation. We measured the survival probability of the ion with respect to the interaction duration. An event is categorized as a survival if the ion remains in the state it was initially prepared in (see Appendix \ref{app:exp-protocol} for details). The data are fitted with an exponential function. The resulting rates are presented in Table \ref{tab:results}, together with those of the theoretical models discussed in the next Sections. A total of 510 and 116 events were observed for Ba$^+$ prepared in $5D_{3/2}$ and $5D_{5/2}$ state, respectively. The contributions of the NRCE, NRQ and FSQ processes are presented as fractions of the total number of counts excluding elastic collisions (EC). The rates are expressed as fractional rates normalized by the experimental Langevin rates $K_{L}^{exp}(5D_{3/2}) = 4.69 \times 10^{-9}$ cm$^{3}$s$^{-1}$ and $K_{L} ^{exp}(5D_{5/2}) = 4.81 \times 10^{-9}$ cm$^{3}$s$^{-1}$ (Appendix \ref{app:langevin}). 

\begin{table*}[!t]\footnotesize
\setlength{\tabcolsep}{0.05mm}
\renewcommand{\arraystretch}{1.7} \addtolength{\tabcolsep}{3 pt}
\begin{center}
\caption{Experimental and theoretical rates for repeated sequences of collision events of a single Ba$^+$ ion prepared in the $5D_{3/2}$ or $5D_{5/2}$ state with ground state Li atoms. The total number of counts is displayed for elastic collisions (EC) and for each process (NRCE, NRQ,FSQ), as well as their ratio (\%) with respect to the number of counts excluding EC. They are converted as experimental rates per Langevin collision ($K_{exp}/K_{L}^{exp}$). The statistical error ($stat$) arises from the fit, whereas the systematic error ($syst$) originates from the density uncertainty of the Li cloud. The theoretical non-thermalized reaction rates per Langevin collision $K_{MCQS}/K_L^{th}$ are computed for a collisional (centre-of-mass) energy $E_i$ expressed as $T_{\mathrm{eff}}=E_{i}/k_B$ of 30~$\mu$K. We estimate a rate $K_{MCQS-L}$ from a Langevin average of $K_{MCQS}$ (see Section \ref{sec:FCQS-MCQS}), thus yielding a range of acceptable theoretical values displayed in the last column. }
\begin{tabular}{cccccc}
\hline \hline
Event&Process&Counts&Ratio($\%$)& $K_{exp.}/K_{L}^{exp}$&  

$K_{MCQS}/K_L^{th}$ ;$\,K_{MCQS-L}/K_L^{th}$
\\
\hline
\multicolumn{6}{l}{$5D_{3/2}$: $K_{L} ^{exp} = 4.69 \times 10^{-9}$ cm$^{3}/s$; $K_{L} ^{th} = 4.81\times 10^{-9}$ cm$^{3}/s$} \\
     &EC  & 177 &-&-&-\\  
Hot  &NRQ &302&90.6(16)&0.154(45)$_{stat}$(27)$_{sys}$& 
0.21;0.18\\   
Loss &NRCE& 31& 9.4(16)&0.016(5)$_{stat}$(3)$_{sys}$& 
0.021;0.012\\   
Total&&510&&&\\ \hline
\multicolumn{6}{l}{$5D_{5/2}$: $K_{L} ^{exp} = 4.81 \times 10^{-9}$ cm$^{3}/s$; $K_{L} ^{th} = 4.81 \times 10^{-9}$ cm$^{3}/s$}\\
        &EC  &41   &-&-&-\\ 
Cold+Hot&FSQ &42+8 & 66(6)& 0.198(26)$_{stat}$(40)$_{sys}$ & 
1.06;0.725\\     
Loss    &NRCE&25   & 34(6)& 0.102(14)$_{stat}$(20)$_{sys}$ & 
0.052;0.16\\ 
Total&&116&&&\\    
\hline \hline
\end{tabular}
\label{tab:results}
\end{center}
\end{table*}

\section{Electronic structure of the L\lowercase{i}B\lowercase{a}$^+$ system}
\label{sec:structure-LZ}

\textit{Hund's case (a) PECs.} We first calculate the LiBa$^+$ PECs without spin-orbit interaction following the methodology of our previous papers (see \cite{aymar2005,aymar2011,aymar2012} and references therein). Briefly, we represent the Li$^+$ and Ba$^{2+}$ ionic cores by effective core potentials completed by core polarization potentials to account for electronic valence-core correlation. Thus only two valence electrons are considered, and the wave functions are represented using a large gaussian basis set. All relevant parameters are reported in the references above. The two-electron Hamiltonian is expressed in this basis, and a full configuration interaction is performed to yield PECs in the body-fixed (BF) frame up to the tenth dissociation limit Li($2s$)+Ba$^+$($6p$). At large internuclear distances $R$, the PECs dissociating into a ground-state Li atom and the Ba$^+$ ion are extrapolated by the term $-C_4/R^4-C_6/R^6$, where $C_{4}=82.2\, a.u.$ is half the static dipole polarizability of the Li atom, calculated within the present basis representation for consistency sake \cite{deiglmayr2008}. The $C_6$ coefficients used in the calculations can be found in Appendix \ref{app:structure}.

\begin{figure}[]
 \centering
 \includegraphics[width=8.5 cm]{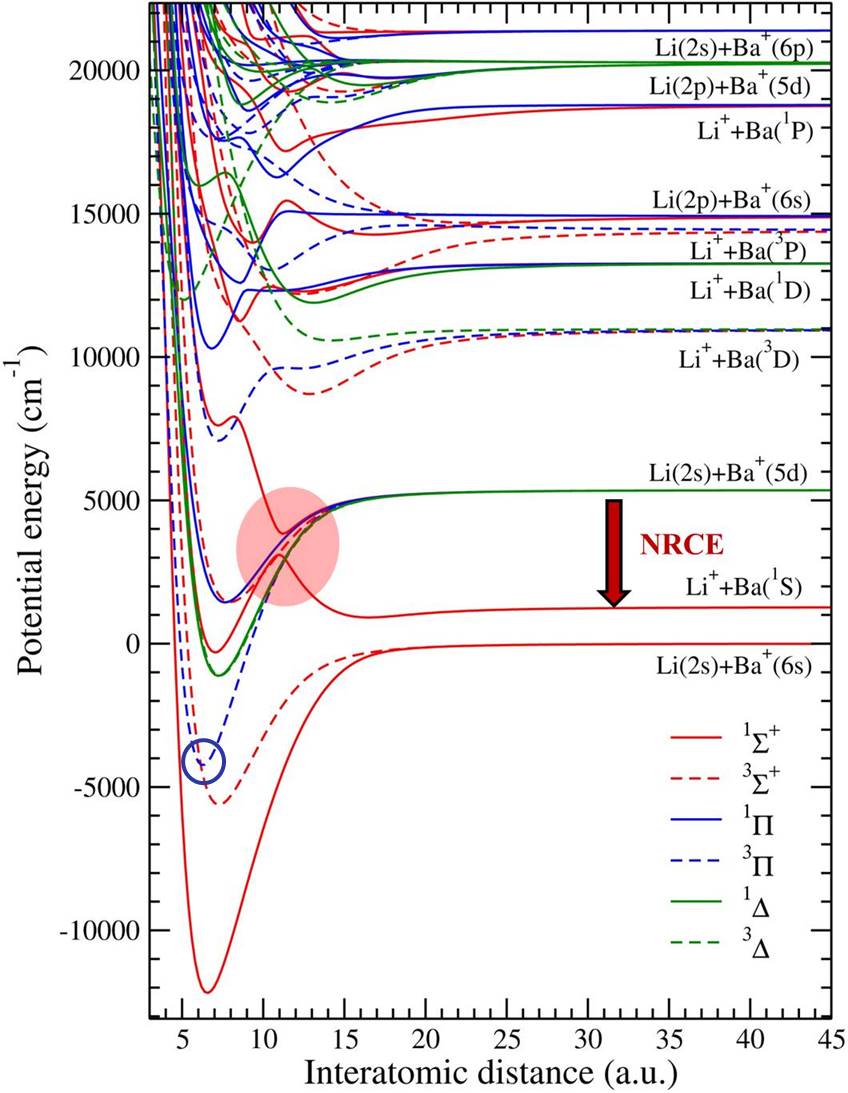}
 \caption{LiBa$^+$ Hund's case (a) PECs in the BF frame up to the Li($2s$)+Ba$^+$($6p$) dissociation limit. The red area locates the $X_1$ avoided crossing between the $2^1\Sigma^+$ and $3^1\Sigma^+$ PECs, inducing NRCE. The blue circle locates the $X_3$ crossing between the $1^3\Sigma^+$ and $1^3\Pi$ PECs.}
 \label{fig:pecs-a}
\end{figure}

The computed Hund's case (a) PECs are displayed in Fig. \ref{fig:pecs-a}, immediately showing that the three lowest dissociation limits of relevance here are quite well isolated from upper ones, in contrast for instance with heavier similar systems like RbBa$^+$ \cite{mohammadi2021} and RbSr$^+$ \cite{aymar2011,benshlomi2020}. Our results are consistent with the recent calculations of LiBa$^+$ electronic structure \cite{weckesser2021b,sardar2023,akkari2024} with a different approach (See Appendix \ref{app:structure} for more details). The most remarkable feature is the avoided crossing (hereafter referred to as the $X_1$ crossing) of the $3\,^1\Sigma^+$ PEC correlated to the Li($2s$)+Ba$^+$($5d$) entrance channel with the $2\,^1\Sigma^+$ PEC around 11~a.u. (1~a.u.=0.052917721092~nm), which will be the main cause of NRCE. It is worth noting that the occurrence of such a crossing in the PECs in the similar entrance channel in the other systems of the same family is not general: the RbBa$^+$ PECs display an avoided crossing in the $^3\Pi$ symmetry \cite{hall2013b,mohammadi2021}, the LiCa$^+$ PECs in the $^3\Sigma^+$ symmetry \cite{saito2017}, the RbSr$^+$ PECs in the $^1\Sigma^+$ and $^3\Sigma^+$ symmetries \cite{benshlomi2020}, and the RbCa$^+$ PECs in the $^1\Sigma^+$, $^3\Sigma^+$, $^1\Pi$ and $^3\Pi$ symmetries \cite{hall2011,hall2013a,xing2022}. This illustrates the variety of the dynamics that can be expected with this class of systems in hybrid traps. Another remarkable feature of the LiBa$^+$ species is the presence of a crossing between the $1\,^3\Sigma^+$ and the $1^3\Pi$ around 6~a.u. (hereafter referred to as the $X_3$ crossing) which is responsible for the additional complexity of the MFRs in the ground state manifold \cite{weckesser2021b}.

\begin{figure}[]
 \centering
 \includegraphics[width=8.8 cm]{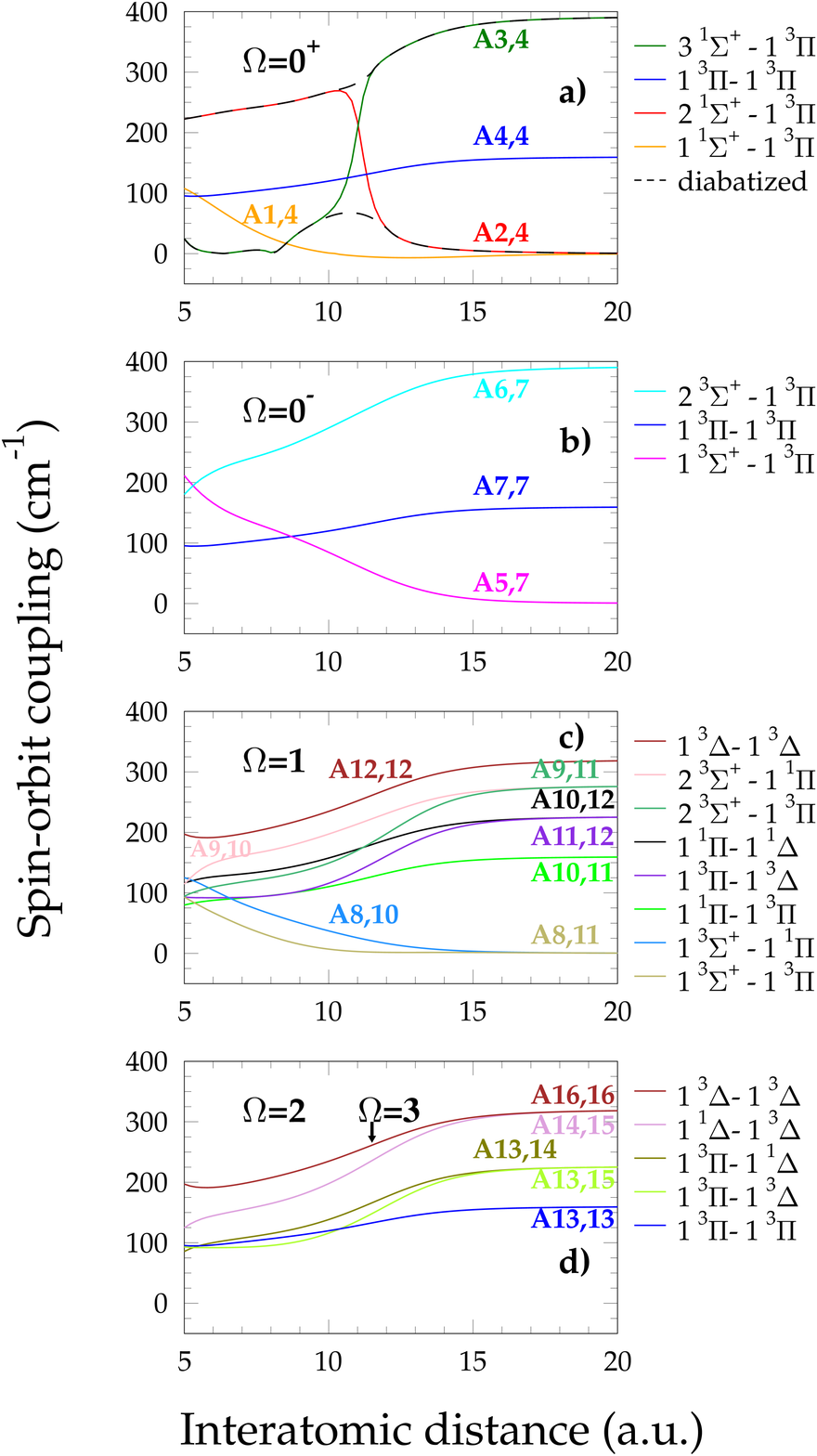}
 \caption{Computed $R$-dependent SOCs (using notations of Table \ref{table:Hammatrix}) between the states correlated to the three lowest LiBa$^+$ dissociation limits, for a) $\Omega=0^{+}$, b) $\Omega=0^{-}$, c) $\Omega=1$, d) $\Omega=2,3$. In panel (a), the diabatized coupling resulting from the linearization of the $X_1$ avoided crossing is drawn as a dashed black line.}
 \label{fig:socs}
\end{figure}

\textit{Spin-orbit couplings.} The $R$-dependent SOCs are obtained following the same quasidiabatic approach extensively described in our previous paper on RbCa$^+$ \cite{xing2022}, inspired by earlier works \cite{cimiraglia1985,angeli1996}. We recall here the main steps for convenience.  A set of reference basis vectors $[|R_1>...|R_N>]$ ($N=30$ here) is defined as the eigenvectors of the electronic Hamiltonian at large internuclear distance (60~a.u.), thus yielding a representation of the separated atom states. A unitary transformation is then applied to express the $N$ lowest adiabatic states $[|\Psi_1^0>...|\Psi_N^0>]$ at arbitrary $R$ on the reference basis set $[|R_1>...|R_N>]$, leading to a quasidiabatic electronic Hamiltonian represented in the separated atom basis. The atomic SOCs are then added to it for the 7 lowest asymptotes, up to Li($2p$)+Ba$^+$($6s$), opening two options: either a full diagonalization of this Hamiltonian to obtain adiabatic PECs including spin-orbit, or to perform the inverse unitary transformation to retrieve SOCs between Hund's case (a) adiabatic states. The latter are displayed in Fig. \ref{fig:socs} for the three lowest dissociation limits, for the Hund's case symmetries labelled with the projection on the molecular axis of the total electronic angular momentum $\Omega=0^{+/-}, 1, 2, 3$. They are labeled according to the notations of the matrix elements of the potential energy matrix reported in Table \ref{table:Hammatrix}. As expected, the couplings between states correlated to the Li($2s$)+Ba$^+$($5d$) asymptote converge toward the relevant atomic values (the atomic spin-orbit splitting between Ba$^+$(5$^2D_{5/2}$) and Ba$^+$(5$^2D_{3/2}$) is 800.955~cm$^{-1}$), while those couplings for states correlated to different asymptotes vanish at large distances.

\begin{table*}[t]
\setlength{\tabcolsep}{0.05mm}
\renewcommand{\arraystretch}{1.5} \addtolength{\tabcolsep}{3 pt}
 \caption{Schematic view of the 16$\times$16 potential energy symmetric matrix in the BF frame, involving the Hund's case (a) molecular states correlated to the three lowest LiBa$^+$ dissociation limits Li($2s$)+Ba$^+$(6s), Li$^+$+Ba($6s^2\,^1S$), Li($2s$)+Ba$^+$(5d), denoted S+S, Ion+S, and S+D, respectively. All blank cells corresponds to zero matrix elements. The G$_{2,3}\equiv$G$_{2,3}$ elements refers to the gaussian coupling associated to the $X_1$ avoided crossing. The A$_{8,11}\equiv$A$_{11,8}$ is the coupling associated with the $X_3$ crossing.}
\begin{center}
 \includegraphics[width=18 cm]{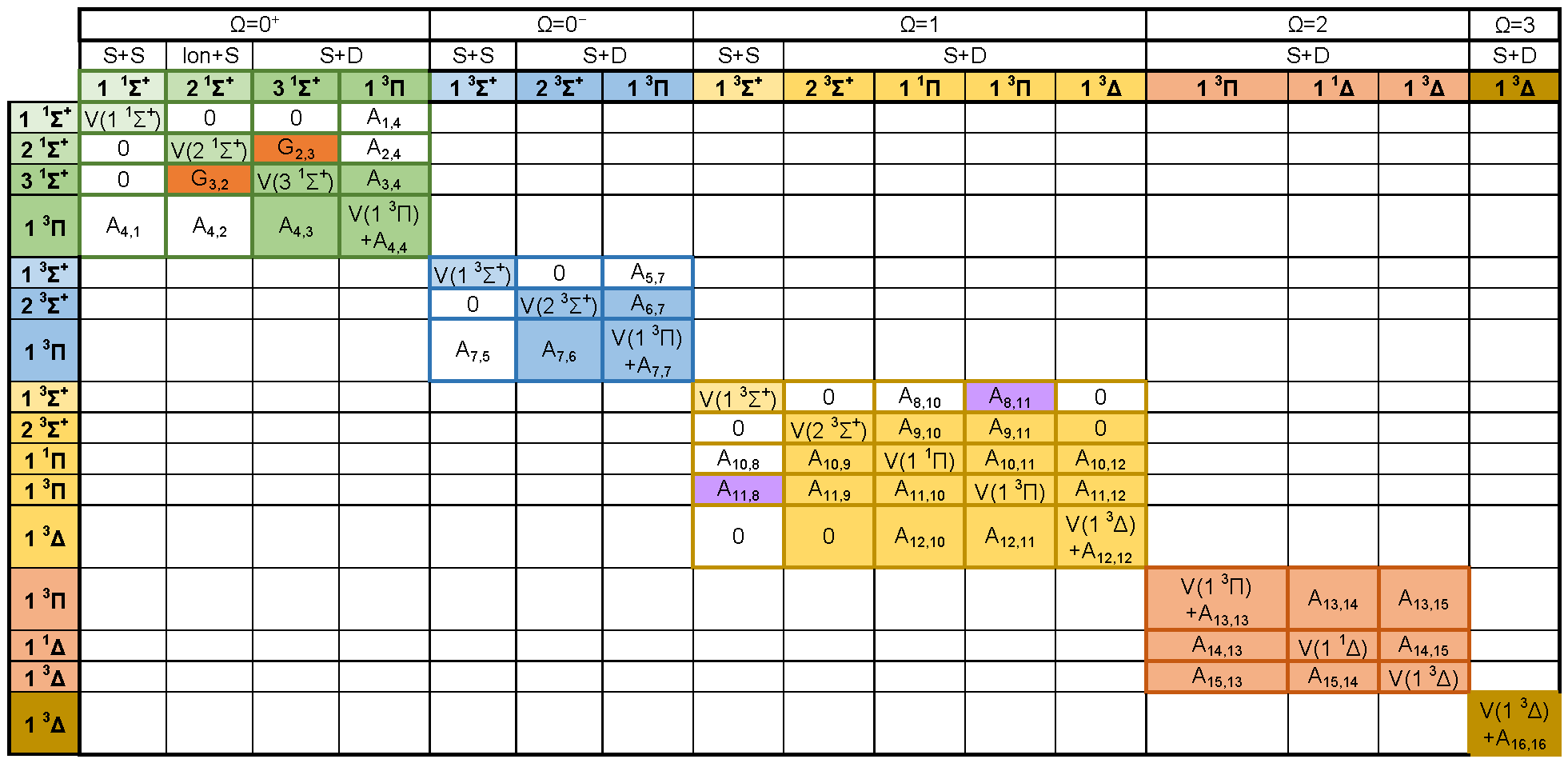}
 \end{center}
 \label{table:Hammatrix}
 \end{table*}

\textit{Landau-Zener modelling.} The $X_1$ avoided crossing requires attention prior to the scattering calculations, as it induces NRCE. We first check its efficiency using a simple Landau-Zener model \cite{landau1967} (see Appendix \ref{app:LZs}, linearizing the avoided crossing at $R_{X1}=11.06$~a.u. where the PECs are split by $2 W_{X1}=0.00359$~a.u. (or 781.7~cm$^{-1}$). We obtain a single-path probability $P_{LZ}=0.769$ and a double-path probability $2P_{LZ}(1-P_{LZ})=0.355$, suggesting that it is quite efficient. At this level of the theory, assuming a statistical population of the initial states, the NRCE probability amounts to $P_{LZ}^{NRCE}=0.355/20=0.0177$, clearly far too small compared to the observations (Table \ref{tab:results}). But this linearization of the $X_1$ crossing will be used in the following to model the interaction around $R_{X1}$ (see Table \ref{table:Hammatrix}) using a gaussian expression $G=W_{X1} exp(-(R-R_{X1})^2/2\delta^2)$, with $W_{X1}=0.001795$~a.u., and a full width $\Gamma=2\sqrt{2ln(2)}\delta$ with $\delta=0.75$~a.u.. We checked the sensitivity of the calculations of the next sections with the empirically chosen width by varying it as $\delta =0.75 \pm 0.5$~a.u., and did not observed a significant effect. The $W_{X1}$ parameter, which is well defined by the PECs, is the main parameter.

We can introduce FSQ in such a simple model by considering the $\Omega=0^+$ block in Table \ref{table:Hammatrix}, diagonalizing it, and setting up a multicrossing LZ model (see Appendix \ref{app:LZs}). An important issue raises here. The marked $X_1$ avoided crossing indicates that the two involved $^1\Sigma^+$ states quite abruptly exchange their electronic character in this region. As we linearized the $X_1$ crossing around $R_{X1}$ by introducing the G$_{2,3}\equiv$G$_{3,2}$ gaussian coupling, we must take into account this diabatization in the SOCs coupling. This is illustrated in Fig. \ref{fig:socs}a: a marked inversion between $A_{2,4}$ and $A_{3,4}$ coupling reflects the presence of the $X_1$ avoided crossing in the related PECs, so that we diabatized these couplings by smoothly joining their left and right branches along the black dashed lines. However the computed probabilities are still in disagreement with the experimental data, due to the statistical assumption for the population of the initial states which weakens the transition probabilities. Note that this was the conclusion of the analysis of Rb+Sr$^+$ collisions reported in \cite{benshlomi2020}, so that quantum scattering calculations must be performed.

\section{Quantum scattering models}
\label{sec:FCQS-MCQS}

In this section we develop the scattering methodology independent of the Li hyperfine level. Details are provided in Appendix \ref{app:cc}.

\textit{Four-channel quantum scattering (FCQS) model.} We set up the FCQS model by extending the previous semiclassical four-channel model. We solve coupled equations in this $\Omega=0^+$ subspace in the space-fixed (SF) frame, but first neglecting the coupling between the internal angular momenta of the atoms and their relative motional angular momentum with momentum $\ell$, referred to as partial wave in the following (see Appendix \ref{app:cc}).

The total cross section for an initial collision energy $E_i=\hbar^2 k_i^2/2\mu$ in the entrance channel $i$ towards the final state $f$ is extracted from the off-diagonal elements of the $\mathbf{S_{\ell}}$ matrix, resulting in a sum over partial waves $\ell$,
\begin{equation}
\begin{aligned}
\sigma(f \gets i,E_i)= \dfrac{\pi}{k_i^2} \sum_{\ell}(2\ell+1)|S_{\ell}(f \gets i)|^2,
\end{aligned}
\label{eq:xsection_ell}
\end{equation}
where $\mu= 10\,481.62$~a.u. is the LiBa$^+$ reduced mass. 

\begin{figure}[]
 \centering
 \includegraphics[width=8 cm]{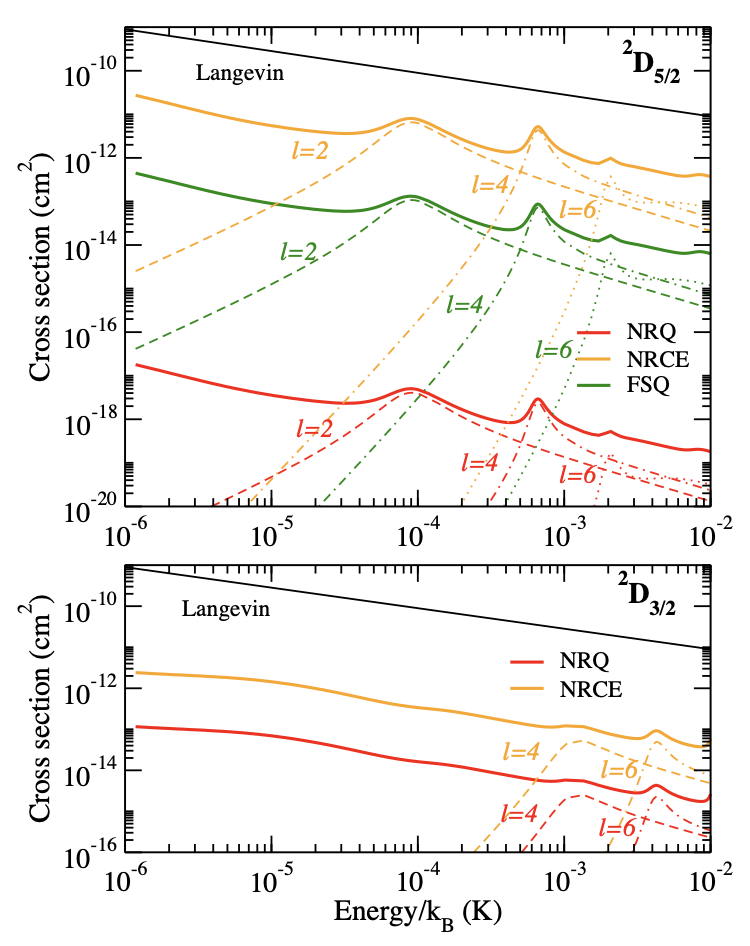}
 \caption{ Non-thermalized total cross sections (full lines) computed with the FCQS model as a function of a fixed initial energy expressed as a temperature $E_i/k_B$, when Ba$^+$ is prepared either in the $5^2D_{5/2}$ and $5^2D_{3/2}$ state. The partial cross sections for the partial waves inducing shape resonances are drawn with dashed lines. The classical Langevin cross section is displayed as a straight line in this double logarithmic scale. }
 \label{fig:xsec-fcqs}
\end{figure}

The calculated partial and total cross sections are presented in Fig. \ref{fig:xsec-fcqs} for the case of the Ba$^+$ ion prepared either in the $5^2D_{5/2}$ or $5^2D_{3/2}$ state. In both cases, the total cross sections for each process exhibit shape resonances in the entrance channel, every two partial waves $\ell$ in accordance with the quantum defect asymptotic theory developed in \cite{gao2013}. As expected, the Langevin cross section $\sigma_L=2\pi C_4^{1/2} E_i^{-1/2}$ resulting from the classical capture model \cite{langevin1905} appears as an upper limit for the quantum cross sections. For the $5^2D_{5/2}$ preparation, the NRQ cross section is negligible compared to the NRCE and FSQ ones, consistently with the experimental observations (Table \ref{tab:results}). But the NRCE is found dominant, in contrast with experiment. A similar conclusion is drawn for the $5^2D_{3/2}$ preparation. Evidently, this quantum scattering approach does not overcome the limitation of the semiclassical LZ models to yield precise cross sections. 

Despite the apparent simplicity of the LiBa$^+$ structure with a single avoided crossing ($X_1$) the coupling of the internal angular momenta with the rotational angular momentum $\ell$ must be taken into account, as already anticipated in our treatment of Rb-Sr$^+$ collisions \cite{benshlomi2020} (which was involving a more complex structure with two avoided crossings).

\textit{Multichannel quantum scattering (MCQS) model}. We consider the $16 \times 16$ potential energy matrix of Table \ref{table:Hammatrix}. We first define the total angular momentum $\vec{J}=\vec{j_{\text{Li}}}+\vec{j_{\text{Ba}}}+\vec{\ell}\equiv \vec{j}+\vec{\ell}$ (with the associated quantum numbers $J$, $j$, $\ell$), its projection $M$ over a quantization axis in the SF frame, and the total parity $p$. As we do not consider any external field, the $M$ quantum number will be omitted in the following. The related frame transformation between the Hund's case (a) molecular basis and the Hund's case (e) basis expressed in the SF frame is described in Appendix \ref{app:atoe}. The relevant quantum numbers for S+S, Ion+S and S+D dissociation limits are reported in Appendix \ref{app:atoe}.

\begin{figure}[]
 \centering
 \includegraphics[width=8 cm]{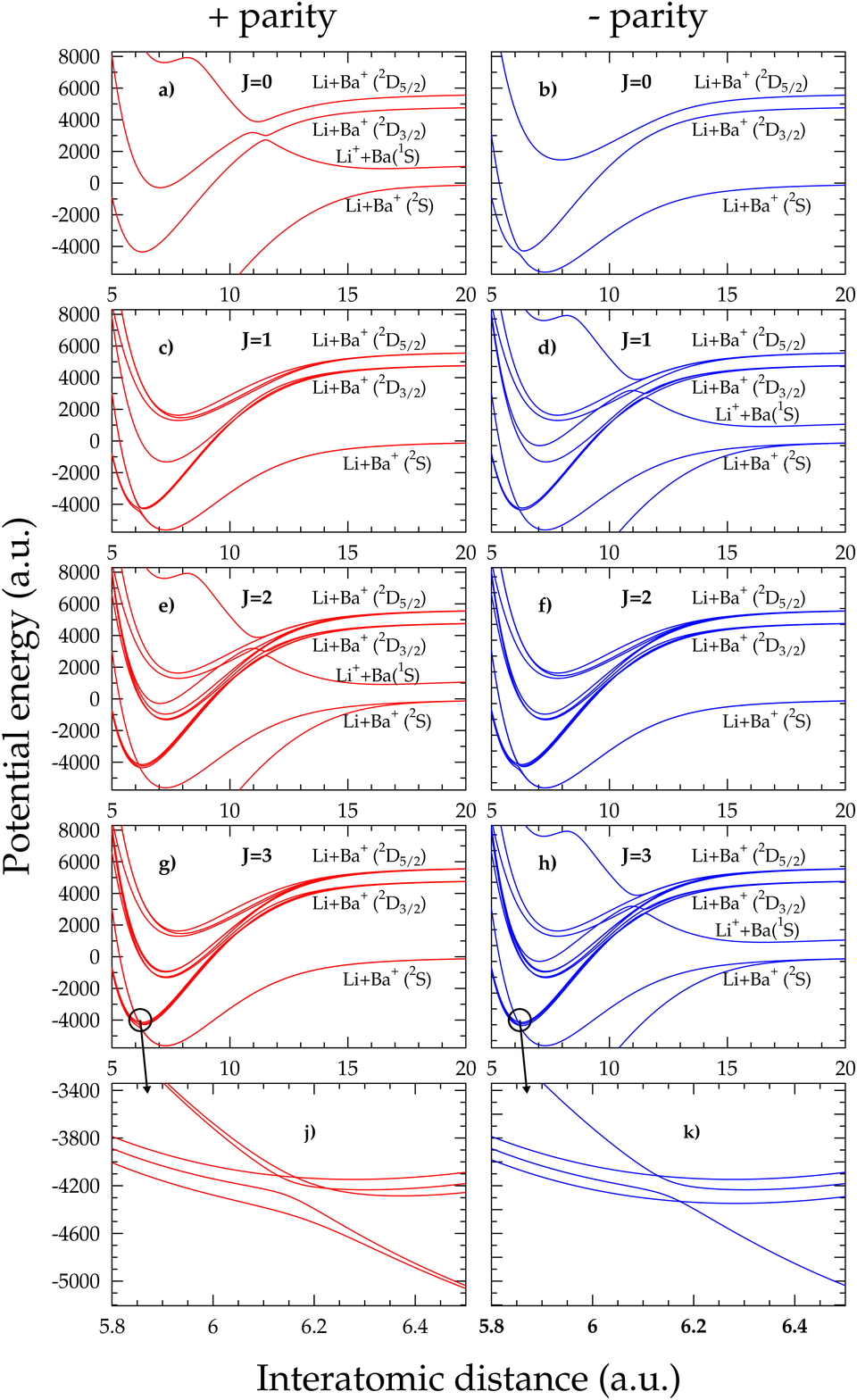}
 \caption{Hund's case (e) potential energy curves for the three dissociation limits S+S, Ion+S, S+D, for $+$ and $-$ parity states. For $J$= 0, 1, 2, 3 and $+$ (resp. $-$) parity, the number of channels is 4, 8, 12, 12 (resp.3, 9, 11, 13). The maximal number of channels is reached for $J \geq 3$: it is 12 (resp. 13) for (odd $J$, $+$) and (even $J$, $-$) (resp. (even $J$, $+$) and (odd $J$, $-$)). Panels j) and k) are zoomed PECs around the avoided crossing $X_3$ marked by black circles on panels g) and h), responsible for NRQ. }
 \label{fig:pecs-e}
\end{figure}

\textit{Hund's case (e) PECs.} They are obtained after diagonalizing the potential energy matrix including rotation for a given $J$ and parity $p$. The results are reported in Fig. \ref{fig:pecs-e} for $J=0-3$ as representative examples. Indeed, due to available angular momenta (see Appendix \ref{app:atoe}), a stable number of channels (12 or 13, depending on the chosen ($J,p$) combination), is reached for $J \geq 3$. For instance, the ($J=0,p=+$) pair only involves the $\Omega=0^+$ subspace and $\ell=2$ (Fig \ref{fig:pecs-e}a). This difference in the maximal number of channels comes from the contribution of the unique $\Omega=0^+$ correlated to the $Ion+S$ channel, which is present in the (even $J,p=+$) and (odd $J,p=-$) cases, and not for the (odd $J,p=+$) and (even $J,p=-$) cases. At short distances, the Hund's case (e) PECs are very similar to Hund's case (c) PECs, but display more complex structure due to the presence of the rotational (Coriolis) coupling. The $X_1$ crossing is still prominent, while exhibiting more complex patterns depending on $J$. One important feature is now that the $X_3$ crossing at 6.2~a.u. now features an avoided crossing (Fig \ref{fig:pecs-e}j,k), due to the combined effect of the rotational coupling, and the indirect SOC (the term $A_{8,11}$ in Table \ref{table:Hammatrix}). Therefore we show strong evidence that NRQ is likely to occur in the present experiment. This crossing has been considered in \cite{weckesser2021b} for the modeling of observed MFRs in LiBa$^+$. Note that a similar indirect SOC has been invoked in Rb-Yb$^+$ cold collisions to explain anomalous hyperfine relaxation \cite{tscherbul2016}.

\textit{Cross sections and rates.} After solving the coupled equations in this basis, Appendix \ref{app:cc}, an $\mathbf{S}$ matrix is obtained for every $J$ value and a given parity $p$, considering the initial state $i$ with collision energy $E_i$ and an outgoing channel $f$. It yields the cross section for a given $J$ and $p$
\begin{equation}
\sigma(E_i,J,p;f)= \dfrac{\pi}{k_i^2}  \sum_{l_i,j_i}\sum_{l_f,j_f}|S(Jl_fj_fp \gets Jl_ij_ip)|^2,
\label{eq:xsection-Jp}
\end{equation}
and then parity-dependent cross section
\begin{equation}
\sigma(E_i,p;f) = \sum_{J}(2J+1) \sigma(E_i,J,p;f), \\
\label{eq:xsection-p}
\end{equation}
and finally total cross section
\begin{equation}
\sigma(E_i;f) = (\sigma(E_i,+1;f)+\sigma(E_i,-1;f))/2.
\label{eq:xsection-tot}
\end{equation}
The $|J,l_i,j_i,p>$ and $|J,l_f,j_f,p>$ vectors represent the chosen initial incoming channels and the allowed final outgoing channels labelled with their quantum numbers valid at infinite distances. The non-thermalized reaction rate is expressed as $K(E_i;f)= (2E/\mu)^{1/2} \sigma(E_i;f)$. We define a thermalized reaction rate at the temperature $T_{\mathrm{eff}}$ assuming a Maxwell-Boltzmann distribution of relative velocities
\begin{equation}
\begin{aligned}
K(T_{\mathrm{eff}};f)&=\dfrac{2}{\sqrt{\pi}(k_BT_{\mathrm{eff}})^{3/2}}\\ &\times\int_0^\infty K(E;f)\sqrt{E}e^{-E/k_BT_{\mathrm{eff}}}dE
\end{aligned}
\label{eq:MBrates}
\end{equation}
The effective temperature in the centre-of-mass \cite{li2020} $T_{\mathrm{eff}}=(m_{Li}T_{Ba^+}+m_{Ba^+}T_{Li})/(m_{Li}+m_{Ba^+})$ is determined by the individual  temperatures $T_{Ba^+} \approx 600\, \mu$K and $T_{Li}\approx 3\, \mu$K \cite{weckesser2021b}, yielding $T_{\mathrm{eff}} \approx 30\,\mu$K.   

\begin{figure}[!h]
 \centering
 \includegraphics[width=8 cm]{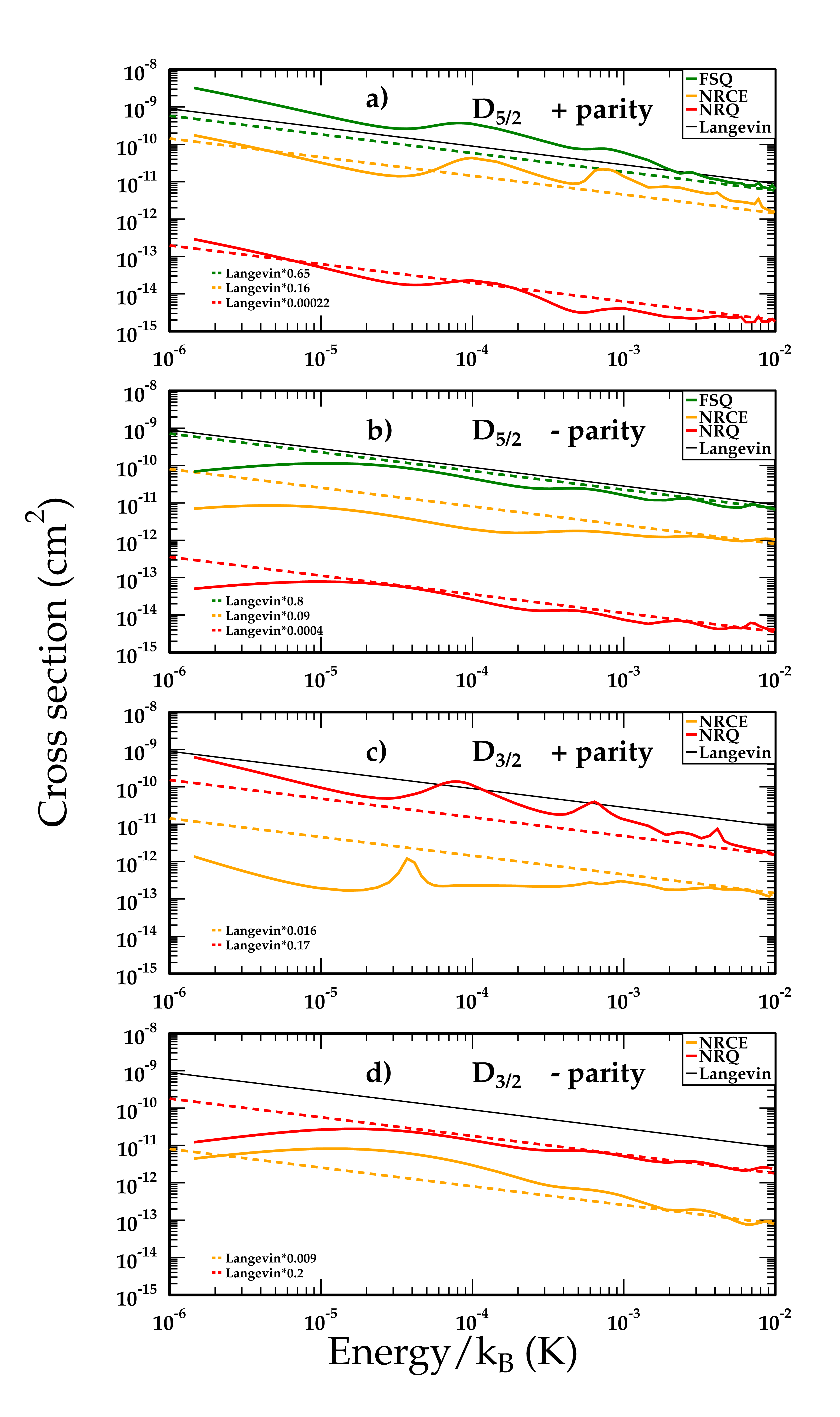}
 \caption{The parity dependent cross sections (Eq. \ref{eq:xsection-p}) as functions of the collision energy (expressed in K), for each allowed process starting from Li($2S_{1/2}$)+Ba$^+$($5D_{5/2}$)  (panels a) and b) and from Li($2S_{1/2}$)+Ba$^+$($5D_{3/2}$) (panels c) and d)). The Langevin cross section $\sigma_L$ is displayed (solid black line), as well as scaled Langevin cross sections (dashed colored lines) for each process. The relative contributions of the various processes could be assessed with these scaling factors regardless the presence of scattering resonances. For the $5D_{5/2}$ case, we find $\sigma_L$(FSQ) $\approx$ 4$\sigma_L$(NRCE) $\approx$ 3000$\sigma_L$(NRQ) for $+$ parity, and $\sigma_L$(FSQ) $\approx$ 8$\sigma_L$(NRCE) $\approx$ 2000$\sigma_L$(NRQ) for $-$ parity.
 Panel c) and d) are for the incoming channels. For the $5D_{3/2}$ case, we find $\sigma_L$(NRQ) $\approx$ 10$\sigma_L$(NRCE) for $+$ parity, and $\sigma_L$(NRQ) $\approx$ 22$\sigma_L$(NRCE) for $-$ parity.}
 \label{fig:xsection52}
\end{figure}

It is worth examining the computed cross sections depending on the parity for FSQ, NRCE and NRQ processes, reported in Fig. \ref{fig:xsection52}. The classical Langevin cross section $\sigma_L$ is reported, as well as partial Langevin cross sections (\textit{i.e.} for each process) estimated by scaling down $\sigma_L$ to adjust it to the computed cross sections for energies above $k_B \times 10$~mK where they are expected to be classical, thus behaving as $E_i^{-1/2}$. The cross sections locally exceeds the Langevin rate due to quantum shape resonances associated to specific $J$ values, or partial waves (See Appendix \ref{app:xsection-J} for more insight). The energy location of these resonances is strongly dependent on the molecular data used in the model: they cannot be predicted, and they have to be detected in the experiment. However the model suggests that such resonances contribute to the dynamics at ultralow energies.

Both parity cases exhibit the same dominant process, namely FSQ and NRQ for Ba$^+$ prepared in the $5D_{5/2}$ and $5D_{3/2}$ state, respectively, which is consistent with the observations. This reveals the strong difference in the structure of the corresponding incoming channels induced by the complex interplay of the various couplings. The NRQ process is found negligible in the $5^2D_{5/2}$ case, in stark contrast with the $5D_{3/2}$ case. This confirms the key importance of the $X_3$ crossing involving PECs correlated to the Li($2S_{1/2}$)+Ba$^+$($5D_{3/2}$) and the Li($2S_{1/2}$)+Ba$^+$($6S_{1/2}$), see Fig. \ref{fig:pecs-e}j,k.

\begin{center}
\begin{figure}[]
 \centering
 \includegraphics[width=8.5 cm] {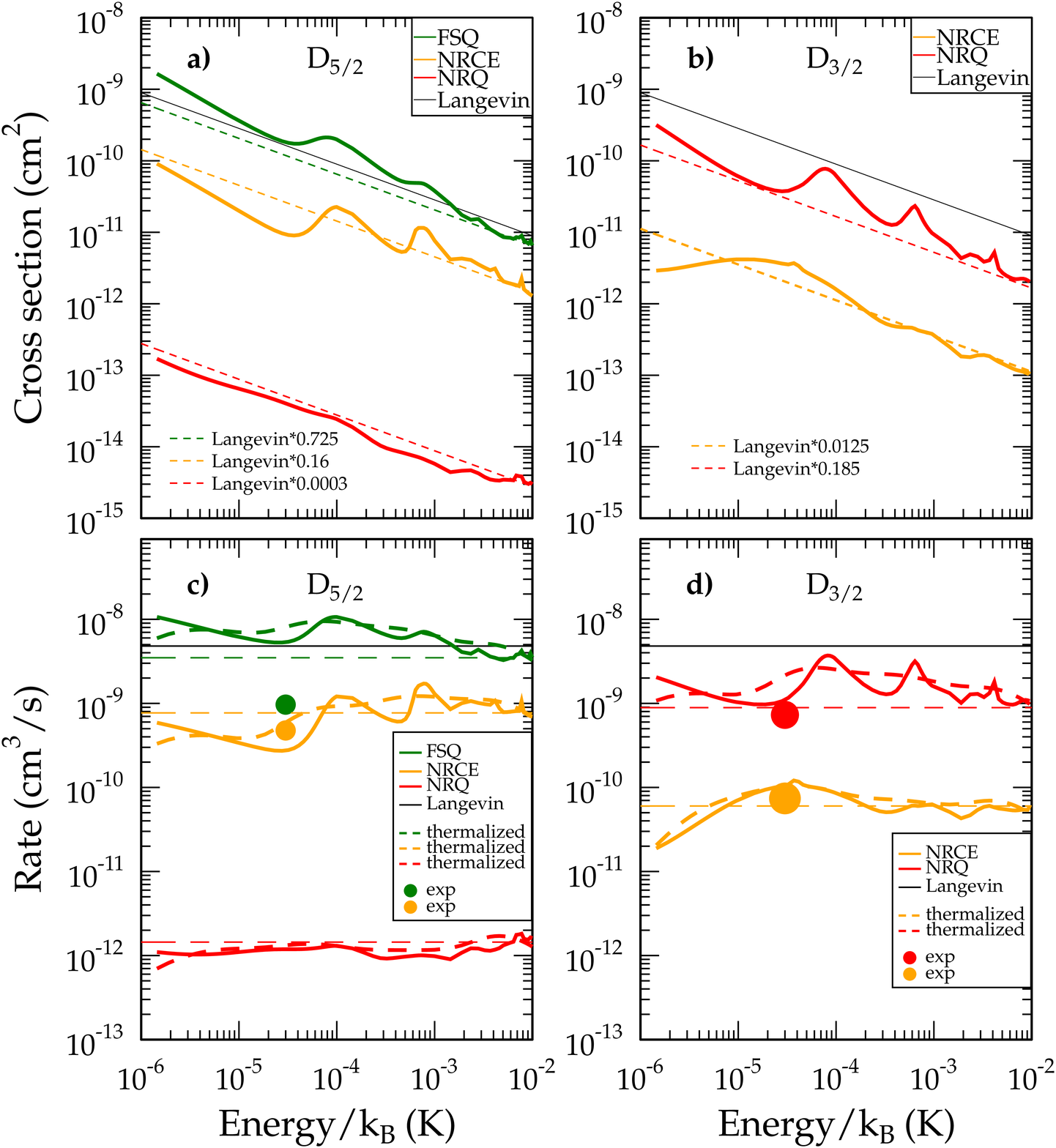}
 \caption{Computed total cross sections for the Li($2S_{1/2}$)+Ba$^+$($5D_{5/2}$) (panel a) and Li($2S_{1/2}$)+Ba$^+$($5D_{3/2}$) (panel b) entrance channels. The Langevin cross section $\sigma_L$ is displayed (black line), as well as scaled Langevin cross sections ( dashed colored lines) for each process. Panels c) and d) show the corresponding rates (colored full lines), the thermalized rates (thick colored long-dashed lines), the scaled Langevin rates (thin colored dashed lines) with the same factors than the cross sections, and the Langevin rate (black line). The circles are the experimental data of Table \ref{tab:results} for $T_{\mathrm{eff}}=30\,\mu$K, with a size consistent with the statistical and systematic errors.}
 \label{fig:xsection-rate}
\end{figure}
\end{center}

The computed total cross sections $\sigma(E_i;f)$ (Eq. \ref{eq:xsection-tot}) are presented in Fig. \ref{fig:xsection-rate}a,b for the $5D_{5/2}$ and $5D_{3/2}$ cases. They obviously display a similar hierarchy between the processes than the parity dependent ones, while the shape resonances are still apparent. They are converted into rates $K(E_i;f)$, and thermally averaged rates $K(T_{\mathrm{eff}};f)$ (Eq. \ref{eq:MBrates}) showing that all resonances are smoothed out (Fig. \ref{fig:xsection-rate} c,d). The corresponding numerical values are reported in Table \ref{tab:results} for the experimental $T_{\mathrm{eff}}=30\,\mu$K \cite{weckesser2021b}, normalized to the theoretical Langevin rate $K_L^{th}=\sigma_L \times (2k_BT_{\mathrm{eff}}/\mu)^{1/2}$ for appropriate comparison with experimental values.

The theoretical rates are found in remarkable agreement with the measured ones around $T_{\mathrm{eff}}=30\,\mu$K, which confirms the strong coupling of the internal angular momenta of the particles with their mutual mechanical rotation all along the collision, as it was anticipated in \cite{benshlomi2020}. We note however the larger discrepancy for the FSQ process in the Li($2S_{1/2}$)+Ba$^+$($5D_{5/2}$) entrance channel, for which the computed rate exceeds the measured one by a factor of about 4. It may be due to the inaccuracy of a given SOC resulting from our quasidiabatic method. The initial polarization of the Li atoms in the experiment could also contribute to this difference, while it is not obvious while it would contribute only for this specific process. 
 
\section{Discussions and Conclusions}
\label{conclusion}

In this work, we detected the outcome of the ultracold collisions between a single Ba$^+$ ion excited in a metastable state immersed in a Li quantum gas close to quantum degeneracy in a hybrid trap. We probed that the dynamics is not restricted to charge exchange, and that several inelastic processes compete with each other and with charge exchange. We measured the branching ratio of these processes, which are found dependent of the initial preparation of the ion. This reveals the complexity of the underlying dynamics. Using a full quantum scattering approach based on high-level electronic structure calculations, we deciphered the main paths, and computed their relative contribution which are found in remarkable agreement with experimental findings, revealing the quality of the molecular data. Similar investigations could be achieved for the same class of systems and will the topic of future works. It is worth noting the special case of LiYb$^+$, in principle very similar to LiBa$^+$ from the experimental point of view, but which is rather tedious to fully describe theoretical: the open $f$-shell of the excited Yb and Yb$^+$ states necessitate the simultaneous consideration of 16 valence electrons, which is far more complicated than in LiBa$^+$.

As stated in the paper, the spin-polarized state of the Li atoms has not been yet taken into account in our approach. This can be accounted for by enlarging the Hilbert space of Table \ref{tab:Hundsbasis} in Appendix \ref{app:atoe} considering the various projections of the angular momenta in the SF frame as additional good quantum numbers. However we note that the series of experimental results on Rb-Sr$^+$($4D_{3/2,5/2}$) reported in \cite{benshlomi2020} yielded no evidence of such dependence with respect to the mutual orientation of the spins of the two particles.

However, reaching the $s$-wave regime, while still challenging, would open new experimental possibilities which could help testing theoretical data even more precisely, while allowing additional level of control of the dynamics of the excited state dynamics. In particular, only the $+$ parity manifold would contribute to the excited state dynamics. Moreover, if both particles would be polarized in the largest spin state $m_{\text{Li}}=+1/2$ and $m_{\text{Ba}}=+5/2$, thus only the $^3\Delta$ molecular state would contribute, so that the ion-atom pair would be protected against any non-radiative decay process. This level is the one with the highest energy in the Zeeman sublevels manifold, so that no MFR would be observable from such an initial collisional channel.

In a broader perspective, the PECs of Fig. \ref{fig:pecs-a} illustrates that the light mass of the system results in dissociation thresholds which quite well separated with broad energy gaps, in contrast with other systems of the same family like RbSr$^+$. Thus LiBa$^+$ Feshbach molecules which would be created from MFR \cite{weckesser2021b,thielemann2024} could be protected against photodissociation by the lasers of the setup in the 6000cm$^{-1}$-10000cm$^{-1}$ approximate range (or roughly $1\mu$m-$1.7\,\mu$m), allowing for longer time to manipulate them. For instance, the $X_3$ crossing  results in a perturbation of the radial wave function (see for instance \cite{dion2001}) of the Feshbach molecules due to the indirect SOC matrix element A$_{8,11}$ (Table \ref{table:Hammatrix}). The $X_3$ crossing happens to be quite aligned with the bottom of the well of the $1\,^1\Sigma^+$ electronic ground state and of the $2\,^1 \Pi$ state (Fig. \ref{fig:pecs-a}. This could represent a pathway for future two-photon experiment aiming at transferring the Feshbach molecules into the lowest vibrational level of ground state LiBa$^+$ ions.

\section{Acknowledgments}
This project has received funding from the European Research Council (ERC) under the European Union’s Horizon 2020 research and innovation programme (grant number 648330), the Deutsche Forschungsgemeinschaft (DFG, grant number SCHA 973/9-1-3017959) and the Georg H. Endress Foundation. P.W. gratefully acknowledges financial support from the Studienstiftung des deutschen Volkes. F.T. and T.S. acknowledge financial support from the DFG via the RTG DYNCAM 2717. The Hungarian and French teams acknowledge support from the CRNS International Emerging Action (IEA) - ELKH, 2023-2024; Program Hubert Curien ”BALATON” (CampusFranceGrantNo.49848TC)–NKFIH TÉT-FR, 2023-2024 (2021-1.2.4- TÉT-2022-00069).

\appendix

\section{Experimental protocol}
\label{app:exp-protocol}

We start each measurement by deterministically preparing a single \ba{} ion in our Paul trap. Here we apply a combination of laser ablation loading with two-photon ionization, resulting in a finite number of Doppler cooled ions. Transferring the ion Coulomb crystal into a far-detuned optical dipole trap at \SI{532}{\nano\meter} while switching off the confining radio-frequency fields of the Paul trap, we can deterministically shape the Coulomb crystal down to a single ion~\cite{LiBa-ion}. We then use the single ion to compensate our axial and radial stray electric fields down to $\lesssim\SI{3}{\milli\volt\per\meter}$ by lowering the confinement of the ion and nulling any observed displacement~\cite{berkeland1998minimization}.

For the interaction with the atoms we prepare the ion in either the \dth{} or \dfh{} manifold (Fig. \ref{Ba+levels}). We prepare the \dth{} state by first switching off the \dth{} repumper for \SI{50}{\milli\second} while continuing to cool on the $D_1$-line (noted COOLER in Fig. \ref{Ba+levels}). For the \dfh{} state, we make use of the off-resonant scattering of our visible (VIS) optical dipole trap~\cite{LiBa-ion}. Here, while Doppler cooling, the \ba{} ion is illuminated with $\approx\SI{5}{W}$ of \SI{532}{nm} laser light until it is successfully shelved. Once prepared, we then shuttle the ion axially and radially out of the trap center to allow for the preparation of the atomic cloud. Note that both electronic preparation schemes do currently not allow to deterministically prepare a dedicated $m_{\mathrm{f}}$ sublevel.

The \li{} atoms are loaded in a conventional magneto-optical trap (MOT) located at the center of the Paul trap and, after a short compression phase, transferred to the far-detuned crossed optical dipole trap (xODT) operated at \SI{1064}{\nano\meter}. We then evaporatively cool the cloud at $B\approx\SI{345}{G}$ to temperatures of \SIrange{1}{3}{\micro\kelvin}. After evaporation, a \SI{15}{\micro\second} laser pulse resonant with the $\ket{m_\textrm{S}=-1/2, m_\textrm{I}=1}\rightarrow P_{3/2}$ transition polarizes the atomic cloud in the $\ket{m_\textrm{S}=-1/2, m_\textrm{I}=0}$ state, where $m_\textrm{S}$ and $m_\textrm{I}$ are the projection on the magnetic field axis of the electronic and nuclear spin, respectively.
Note that for lower magnetic fields the $\ket{m_\textrm{S}=-1/2, m_\textrm{I}=0}$ state can be expressed as $\ket{f=1/2, m_\textrm{f}=-1/2}$. We then shift the magnetic field to $B = \SI{293}{G}$ where it remains during the interaction phase and the subsequent detection of the atomic cloud. In principle, we can individually align the two xODT beams to the position of the ion with two piezo-controlled mirrors. Overlap between the atomic cloud and the ion is independently verified by measuring the inelastic ion-loss probability for different ion displacements and continuously checked throughout the measurement. 

\begin{figure}[]
 \centering
 \includegraphics[width=8cm]{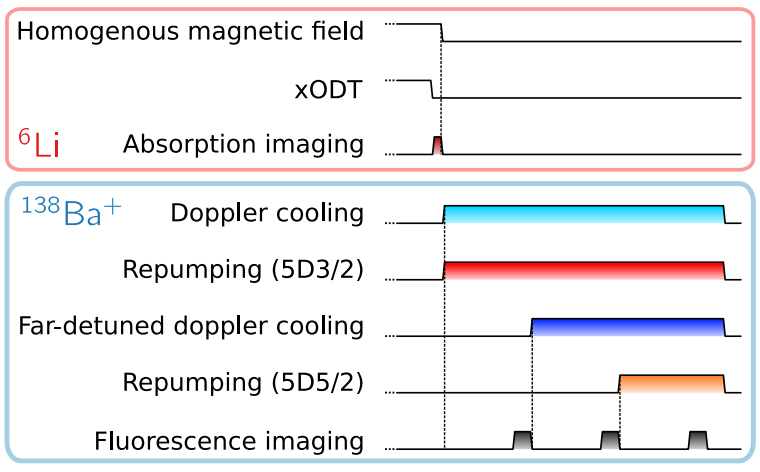}
 \caption{Experimental protocol of the \li{} detection and the subsequent \ba{} product state detection sequence (durations are not to scale). At the end of the atom-ion interaction phase, the xODT is switched off. After a short expansion duration of $100$~$\mu$s we perform high-field absorption imaging of the atomic cloud. We then release the magnetic field to 4~G, and detect and classify the ion's electronic state into four possible outcomes. First, the near-detuned Doppler cooling lasers are switched on for \SI{1}{s}, followed by \SI{300}{ms} of fluorescence imaging. An ion detected in this stage is classified as \emph{cold} \soh~or~\dth. Secondly, we switch on a far-detuned cooling beam, also followed by fluorescence detection. An ion appearing in this stage is classified as \textit{hot}. Lastly, we shine in the \SI{614.9}{nm} rempumper to detect whether the ion is shelved in the \dfh-state. If the ion is not detected after any of the steps, the event is classified as a \emph{loss}.}
 \label{fig:product_state_detection}
\end{figure}

For the interaction we shuttle the ion back to the trap center and let the ion interact with the atomic ensemble for variable time. Afterwards we apply the protocol depicted in Fig.~\ref{fig:product_state_detection} to detect both the atomic ensemble as well as the outgoing \ba{} electronic state. First, we switch off the xODT and after a short time of flight, the atomic cloud is absorption imaged on a closed cycle transition for \SI{15}{\micro\second} at $B = \SI{293}{G}$. The magnetic fields are then ramped down to $B \approx \SI{4}{G}$ for \ba{} state detection, which consists of three phases.

In each phase the ion is illuminated by different detection lasers for \SI{1}{\second}, followed by a \SI{300}{\milli\second} fluorescence image (CCD camera). In the first phase, only the Doppler cooling and \dth-repumper lasers are switched on. This will reveal ions that are either in the \dth-\, or  \soh-state with a temperature below $\approx \SI{50}{\kelvin}$. The latter is limited by the spatial overlap of the cooling beam with the ion. Next we additionally shine in a far-detuned ($\delta\approx15\,\Gamma_\text{nat}$) Doppler cooling laser with larger waist to recool hot ions with a kinetic energy equivalent to several hundred Kelvin. Finally we apply the \dfh-repumper to deshelve ions that are in the \dfh-state after the interaction.  If the ion is not detected during any of the phases, the event is classified as a loss, which we attribute to NRCE.

Because we cannot distinguish between the \soh- and \dth-state, due to the necessity of the \dth-repumper for fluorescence detection, we have to interpret the outcome of the first detection phase depending on the initial state of the ion. If the ion is initially prepared in the \dfh-state, we assume that it has undergone FSQ to the \dth-state, because quenching to the \soh-state would heat the ion by $\approx\SI{280}{K}$, which is too hot for direct fluorescence detection and recooling. Similarly for an ion initially in the \dth-state we assume that it has remained in that state. To obtain the survival probability of an ion in the \dth-(\dfh-)state, we calculate the  relative numbers of ions detected in the first (third) detection phase.

\section{Calibrating the Langevin scattering rate}
\label{app:langevin}

To compare the observed reaction rates to the Langevin rate $K_\text{L}$, we measure the number of atoms $N$, the radial trap frequency $\omega_\text{rad}$, the axial size $\sigma_\text{ax}$ and the temperature $T$ of the atomic cloud to obtain the number density $n = \frac{1}{(2\pi)^{3/2}} \, \frac{m_\text{Li} \,\omega_\text{rad}^2}{k_\text{B} \,T \, \sigma_\text{ax}} \, N$. We adjust the densities for interaction with the ion in the \dth{} or \dfh{} state to $n = \SI{1.3(2)e11}{\centi\meter^{-3}}$ and $\SI{1.6(4)e11}{\centi\meter^{-3}}$ respectively. The Langevin collision rate for atom-ion interactions is $K_\text{L} = 2\pi\,n\,\sqrt{2\,C_4/\mu}$, with the reduced mass $\mu$ and the induced dipole coefficient $C_4$.
The corresponding Langevin rates are $K_\text{L} = \SI{610(100)}{\per\second}$ and \SI{770(180)}{\per\second}.

\section{Correcting imperfect \dfh{} state preparation}
\label{app:d52prep}

When we conduct experiments with the ion initially in the \dfh-state, but without the presence of atoms, we observe a cold ion in the first detection phase in \SI{17.5(14)}{\percent} of all cases. We latter identified this to a leakage of \SI{615}{nm} rempumper light into the chamber. As the interaction duration is orders of magnitude shorter than the preparation of the atomic cloud, we can assume that the ion is pumped to the ground state before the interaction begins. Having observed that the \soh~state is reactionally stable up to \SI{1}{\second} of interaction time at the given densities, we rescale the respective product rate of \dfh~experiments accordingly.

\section{Potentials and spin-orbit couplings}
\label{app:structure}

In Table \ref{table:constants} we present a comparison of the equilibrium distance and the well depth of our computed PECs with those recently reported in our paper \cite{weckesser2021b} and elsewhere \cite{sardar2023,akkari2024} obtained with other computational approaches. While being all consistent with each other, significant dispersion of the results is visible. It is tedious to decide which calculations provide the most accurate predictions, as their accuracy strongly depends on the details of the implementation of each calculation within a given methodology. We recall that our calculations uses a full configuration interaction, in contrast with the other references, which is often an argument in favor of a better accuracy. In contrast, the position of the $X_1$ and $X_3$ crossings are very similar in all methods, as well as the energy separation for $X_1$. This is encouraging as these are the relevant parameters which control the dynamics treated in the present paper.

For completeness we also list the $C_6$ van der Waals coefficient which has been used in addition to the $C_4$ (identical for all PECs but the $2^1\Sigma^+$) coefficient to fit and extrapolate the PECs at large distances.

\begin{table}[tb]
\centering
\caption{equilibrium distance $R_e$ and well depth $D_e$  of the LiBa$^{+}$ PECs. The location and energy of the $X_1$ and $X_3$ crossings are also given. The van der Waals coefficient $C_6$ used to extrapolate the PECs at large distance is displayed in the last column.}
\begin{tabular}{ cccc } 
\hline
 State & $R_e$ (a.u.) & $D_e$ (cm$^{-1}$)&$C_{6}$ (a.u.) \\
\hline
1$^1$$\Sigma^+$ & 6.60 & 12189 & $-14820.26$ \\ 
&6.70\cite{sardar2023}&11627\cite{sardar2023}& \\ 
&6.61\cite{akkari2024}&11846\cite{akkari2024}& \\ 
&6.75\cite{weckesser2021b}&11860\cite{weckesser2021b}& \\ 
1$^3$$\Sigma^+$ & 7.28& 5619 & $-14821.26$ \\ 
&7.56\cite{sardar2023}&4784\cite{sardar2023}& \\ 
&7.41\cite{akkari2024}&5401\cite{akkari2024}& \\ 
&7.46\cite{weckesser2021b}&5178\cite{weckesser2021b}&\\
2$^1$$\Sigma^+$ & 7.02& 1580& $-21744.76$ \\ 
&7.57\cite{sardar2023}&1206\cite{sardar2023}&\\ 
&7.05\cite{akkari2024}&2246\cite{akkari2024}& \\ 
3$^1$$\Sigma^+$ & 11.20& 1524& $-2327.73$ \\ 
&10.90\cite{sardar2023}&1833\cite{sardar2023}& \\ 
&11.41\cite{akkari2024}&1569\cite{akkari2024} & \\ 
2$^3$$\Sigma^+$ & 7.98& 3905& $-2327.73$ \\
&7.48\cite{sardar2023}&5961\cite{sardar2023}& \\
&8.11\cite{akkari2024}&3952\cite{akkari2024}& \\
1$^1$$\Pi$ & 7.60& 3915& $-3180.61$ \\ 
&7.75\cite{akkari2024}&4034\cite{akkari2024}& \\ 
1$^3$$\Pi$ & 6.28 & 9589& $-3180.72$ \\ 
&6.42\cite{sardar2023}&8935\cite{sardar2023}& \\ 
& 6.29\cite{akkari2024}&9390\cite{akkari2024}& \\ 
1$^1$$\Delta$ & 7.29& 6473& $-5686.68$ \\ 
&7.18\cite{sardar2023}&5729\cite{sardar2023}& \\ 
&7.35\cite{akkari2024}&6250\cite{akkari2024}& \\ 
1$^3$$\Delta$ & 7.20& 6477& $-5686.68$ \\ \hline
&7.42\cite{akkari2024}&6235\cite{akkari2024}& \\ \hline
$X_1$  & 11.06& 3450& \\ 
&10.88\cite{sardar2023}&4171\cite{sardar2023}&  \\ 
&11.30\cite{akkari2024}&3465\cite{akkari2024}&  \\ 
$X_3$  & 6.15& -4205& \\ 
&6.02\cite{sardar2023}&-2326\cite{sardar2023}& \\ 
&6.14\cite{akkari2024}&-3973\cite{akkari2024}& \\ 
\hline
\end{tabular}
\label{table:constants}
\end{table}

\section{Landau Zener model}
\label{app:LZs}

The Landau-Zener transition probability \cite{landau1967} between two locally linear PECs crossing in $R_c$ for a single path through a crossing is $P_{LZ}=exp(-2\pi W_c^2/(v_c\Delta F_c))$. The coupling parameter $W_c$ is the energy half-spacing of the two adiabatic PECs in $R_c$, and $\Delta F_c$ is the difference of slopes of the two linearized branches. The relative local velocity of collisions $v_c=\sqrt{2(E_i-U_c)/\mu}$ results from the difference between the initial collision energy $E_i$ and the potential energy $U_c$ in $R_c$, with $\mu$ the reduced mass of the system. In the ultracold regime, $E_i$ is negligible compared to $U_c$. The double-path probability is obtained according to $2P_{LZ}(1-P_{LZ})$.

\begin{figure}[]
 \centering
 \includegraphics[width=8 cm]{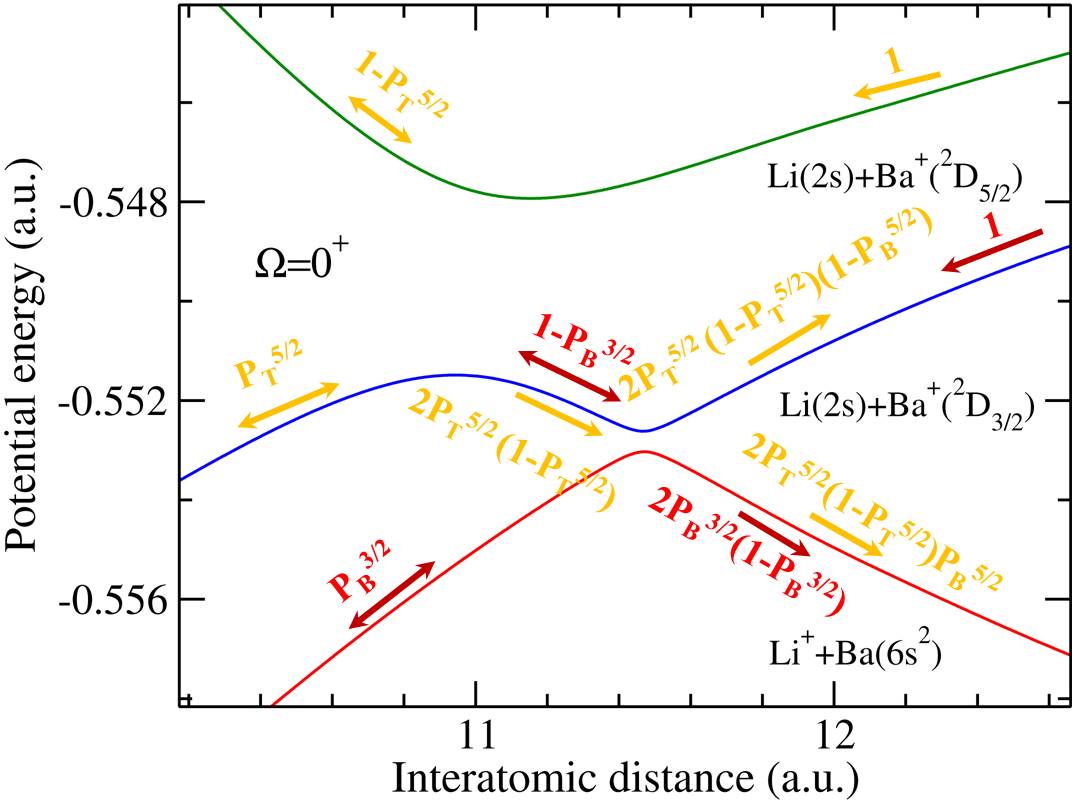}
 \caption{The LiBa$^+$ PECs around the crossing point $R_c$ after the diagonalization of the $\Omega=0^+$ submatrix in Table \ref{table:Hammatrix}. The incoming and outgoing LZ probabilities are marked by red and yellow arrows. A unit probability is assumed in the Li($2S_{1/2}$)+Ba$^+$($5D_{5/2}$) (yellow arrows) and the Li($2S_{1/2}$)+Ba$^+$($5D_{3/2}$) (red arrows) entrance channels.}
  \label{fig:FCLZ}
\end{figure}

For the four-channel LZ (FCLZ) model invoked in the main text, Fig. \ref{fig:FCLZ} displays the shape of the corresponding PECs around $R_{X1}$ after diagonalizing the $\Omega=0^+$ submatrix in Table \ref{table:Hammatrix}, with the corresponding partial probabilities. Labelling with $T$ and $B$ the upper and the lower avoided crossings in the figure, the single-path probabilities entering from the Li($2S_{1/2}$)+Ba$^+$($5D_{5/2}$) or the Li($2S_{1/2}$)+Ba$^+$($5D_{3/2}$) amounts to $P_T^{5/2}=0.264$, $P_B^{5/2}=0.982$ and $P_B^{3/2}=0.979$. Including the statistical weights $1/12$ and $1/8$ for the initial population of the $\Omega=0^+$ state within the $5D_{5/2}$ and $5D_{3/2}$ incoming channels yields $P^{5/2}_{NRCE} =P^{5/2}_T(1-P^{5/2}_T)(1-P^{5/2}_B)/6=0.0318$, $P^{3/2}_{NRCE}=(1-P^{3/2}_B)P^{3/2}_B/4= 0.0051$, and $P^{5/2}_{FSQ}$=$P_T(1-P_T)P^{5/2}_B/6=0.0006$.

\section{Coupled-channel equations for quantum scattering}
\label{app:cc}

We first disregard the coupling between the mechanical rotation between the two particles $\ell$ and the internal angular momenta, so that the total wave function of the colliding pair can be expressed as a partial wave expansion, and each partial wave $\ell$ is treated independently. The corresponding Hamiltonian $\mathbf{H}_{\ell}$ considering the electronic interactions (PECs and SOCs) and the uncoupled rotation of the nuclei, can be described as
\begin{equation}
\begin{aligned}
\mathbf{H}_{\ell}(R)=-\dfrac{\hbar^2}{2\mu}\dfrac{d^2}{d R^2}\mathbf{I}+\dfrac{\hbar^2 l(l+1)}{2\mu R^2}\mathbf{I}+\mathbf{V}(R)+\mathbf{V}_{soc}(R)
\end{aligned}
\end{equation}
where $\mu$ is the LiBa$^+$ reduced mass, $R$ is the internuclear distance,  $\mathbf{V}(R)$ the electronic potential energy matrix, $\mathbf{V}_{soc}(R)$ the spin-orbit matrix, and $\mathbf{I}$ the identity matrix. $\ell$ is the partial wave, i.e. the mechanical rotation of the colliding nuclei in SF frame.  We solve the coupled equations using log derivative method \cite{johnson1973,alexander1987} with a constant step-size 0.005~a.u. for a given collision energy $E$. Since the rotational interaction (varying as $1/R^2$) dominates the electrostatic interaction (varying as $1/R^4$), the scattering wave function is defined by Riccati-Bessel functions at infinity (actually 10 000~a.u. in the computations). We then extract the reaction matrix $\mathbf{K}$ holding close and open channels and the scattering matrix $\mathbf{S}$ containing only open channels.

We apply the same numerical approach for the MCQS calculations in the SF frame using the basis transformation defined in the next Section.

\section{Frame transformation from Hund's case (a) in the BF frame to Hund's case (e) basis in the SF frame}
\label{app:atoe}

To describe the states related to the Li$^+$+Ba and Li+Ba$^+$ dissociation limits, we use the properly symmetrized fully-coupled Hund's case (e) basis functions in the SF frame $|j_a j_b j \ell J M p>$, where the quantum  numbers are associated to the angular momenta $\vec{j}=\vec{j}_a+\vec{j}_b$, $\vec{J}=\vec{j}+\vec{\ell}$, $M$ being the projection of $\vec{J}$ on a quantization axis. The basis functions have a defined parity $p=(-1)^{L_a+L_b+\ell}$, where $\vec{L}_{\mathrm{Li}}$ and $\vec{L}_{\mathrm{Ba}}$ are the electronic angular momenta of the atoms. In the following, we omit $M$ as we do not consider any external field.

In this basis the matrix elements $H_{rot}^{ij}$ are simply equal to $\delta_{ij}\hbar^2\ell_i(\ell_i+1)/2\mu R^2$, where $\ell_i$ denotes the rotational angular momentum in the channel $i$. The BF to SF frame transformation is applied to $V(R)+V_{soc}(R)$. The new $\overline{V(R)}$ is not diagonal and the $\overline{V_{soc}(R)}$ differs from the spin-orbit matrix in the FCLZ model.

In the BF frame, the projection of the total angular momentum $J$ on the molecular axis is $\Omega=\Lambda+\Sigma$, where $\Lambda$ and $\Sigma$ are the projections of the electronic orbital $\vec{L}$ and the spin angular momenta $\vec{S}$ on the molecular axis, respectively. The molecular basis with parity $(-1)^p$ in Hund's case (a) is
\begin{equation}
\begin{aligned}
|\Lambda S \Sigma J p&>=(2-\delta_{\Lambda,0}\delta_{\Sigma,0})^{-1/2} \\
&\times \{|\Lambda S \Sigma J\Omega >+(-1)^{J-S+p} \\
&\times (1-\delta_{\Lambda,0}\delta_{\Sigma,0})|-\Lambda S-\Sigma J-\Omega > \}
\end{aligned}
\end{equation}
The transformation elements from (a) to (e) is obtained by
\begin{equation}
\begin{aligned}
<j_aj_bj\ell Jp&|\Lambda S \Sigma
JMp>=(-1)^{\ell-\Omega-J}(2-\delta_{\Lambda,0}\delta_{\Sigma,0})^{-1/2}\\
&\times [1+(-1)^{L_a+L_b+\ell+p}(1-\delta_{\Lambda,0}\delta_{\Sigma,0})] \\
&\times \sqrt{(2S+1)(2j_a+1)(2j_b+1)} \\
&\times<l0|j-\Omega,J\Omega><L\Lambda|L_a\Lambda_a,L_b\Lambda_b>\\
&\times \left\{\begin{matrix}
L_a&S_a&j_a\\
L_b&S_b&j_b\\
L&S&j\\
\end{matrix}\right\}<j\Omega|L\Lambda,S\Sigma>
\end{aligned}
\end{equation}
where $\vec{L}=\vec{L}_a+\vec{L}_b$, $\Omega=0^{\pm}, 1,2, 3$, and the sharp and curly brackets denote $3j-$ and $9j-$ coefficients, respectively. When squared, these matrix elements determine the weights of Hund's case (a) $|L \Lambda S \Sigma>$ channels in the Hund's case (e) channels. The relevant quantum numbers for S+S, Ion+S and S+D dissociation limits are summarized in Table \ref{tab:Hundsbasis}. We note that in the $|\Omega|=0$ case, the parity of the $^1\Sigma^+_0$ (resp. $^3\Sigma^+_0$) state is $+$ for even (resp. odd) $J$ values. Thus for $+$ parity, $|\Omega|=0^+$ is involved only for even $J$, and $|\Omega|=0^-$ for odd $J$. For $-$ total parity  the situation is reversed, \textit{i.e.} $|\Omega|=0^-$ only for even $J$ values, and $|\Omega|=0^+$ for odd $J$ values.

\begin{table}[]\footnotesize
\renewcommand{\arraystretch}{1.55} \addtolength{\tabcolsep}{3 pt}
\caption{The good quantum numbers for S+S, Ion+S and S+D dissociation limits for Hund's cases (a), (c) and (e), for even and odd $J$ values, thus determining the correspondence with the total parity $+$ and $-$. There is no line-to-line correspondence between the columns.  }
\begin{center}
\begin{tabular}{cccccc}
\hline \hline
\multicolumn{6}{c}{$J$\,-\,even/odd }\\
\multicolumn{2}{c}{(a) ($^S\Lambda_{|\Omega|}$)} &\multicolumn{2}{c}{ (c) ($|\Omega|$)}&\multicolumn{2}{c}{(e) ($j_{Li},j_{Ba},j,l$)} \\
$+/-$&$-/+$&$+/-$&$-/+$&$+/-$&$-/+$\\ \hline
 \multicolumn{6}{c}{S+S} \\
 $^1\Sigma^+_0$&$^3\Sigma^+_0$&$0^+$&$0^-$&
 ($\frac{1}{2}$,$\frac{1}{2}$,0,J)&
 ($\frac{1}{2}$,$\frac{1}{2}$,1,J-1) \\  
 $^3\Sigma^+_1$&$^3\Sigma^+_1$&$1$&$1$&
 ($\frac{1}{2}$,$\frac{1}{2}$,1,J)&($\frac{1}{2}$,$\frac{1}{2}$,1,J+1) \\ 
 \multicolumn{6}{c}{Ion+S} \\
 $^1\Sigma^+_0$&-&$0^+$&-&
 (0,0,0,J)&
 (0,0,0,J) \\  
  \multicolumn{6}{c}{S+D} \\
$^1\Sigma^+_0$&$^3\Sigma^+_0$&$0^+$&$0^-$&
($\frac{1}{2}$,$\frac{3}{2}$,1,J)&($\frac{1}{2}$,$\frac{3}{2}$,1,J-1) \\  
$^3\Sigma^+_1$&$^3\Sigma^+_1$&1&1&
($\frac{1}{2}$,$\frac{3}{2}$,2,J-2)&($\frac{1}{2}$,$\frac{3}{2}$,1,J+1) \\ 
$^1\Pi_1$&$^1\Pi_1$&1&1&
($\frac{1}{2}$,$\frac{3}{2}$,2,J)&($\frac{1}{2}$,$\frac{3}{2}$,2,J-1)\\ 
$^3\Pi_0$&$^3\Pi_0$&$0^+$&$0^-$&
($\frac{1}{2}$,$\frac{3}{2}$,2,J+2)&($\frac{1}{2}$,$\frac{3}{2}$,2,J+1) \\ 
$^3\Pi_1$&$^3\Pi_1$&1&1&
($\frac{1}{2}$,$\frac{5}{2}$,2,J-2)&($\frac{1}{2}$,$\frac{5}{2}$,2,J-1) \\ 
$^3\Pi_2$&$^3\Pi_2$&2&2&
($\frac{1}{2}$,$\frac{5}{2}$,2,J)&($\frac{1}{2}$,$\frac{5}{2}$,2,J+1) \\ 
$^1\Delta_2$&$^1\Delta_2$&2&2&
($\frac{1}{2}$,$\frac{5}{2}$,2,J+2)&($\frac{1}{2}$,$\frac{5}{2}$,3,J-3) \\ 
$^3\Delta_1$&$^3\Delta_1$&1&1&
($\frac{1}{2}$,$\frac{5}{2}$,3,J-2)&($\frac{1}{2}$,$\frac{5}{2}$,3,J-1) \\ 
$^3\Delta_2$&$^3\Delta_2$&2&2&
($\frac{1}{2}$,$\frac{5}{2}$,3,J)&($\frac{1}{2}$,$\frac{5}{2}$,3,J+1) \\ 
$^3\Delta_3$&$^3\Delta_3$&3&3&
($\frac{1}{2}$,$\frac{5}{2}$,3,J+2)&($\frac{1}{2}$,$\frac{5}{2}$,3,J+3) \\     
\hline \hline
\end{tabular}
\end{center}
\label{tab:Hundsbasis}
\end{table}

\section{Partial waves and shape resonances in the cross sections}
\label{app:xsection-J}

\begin{figure}[]
 \centering
 \includegraphics[width=8 cm]{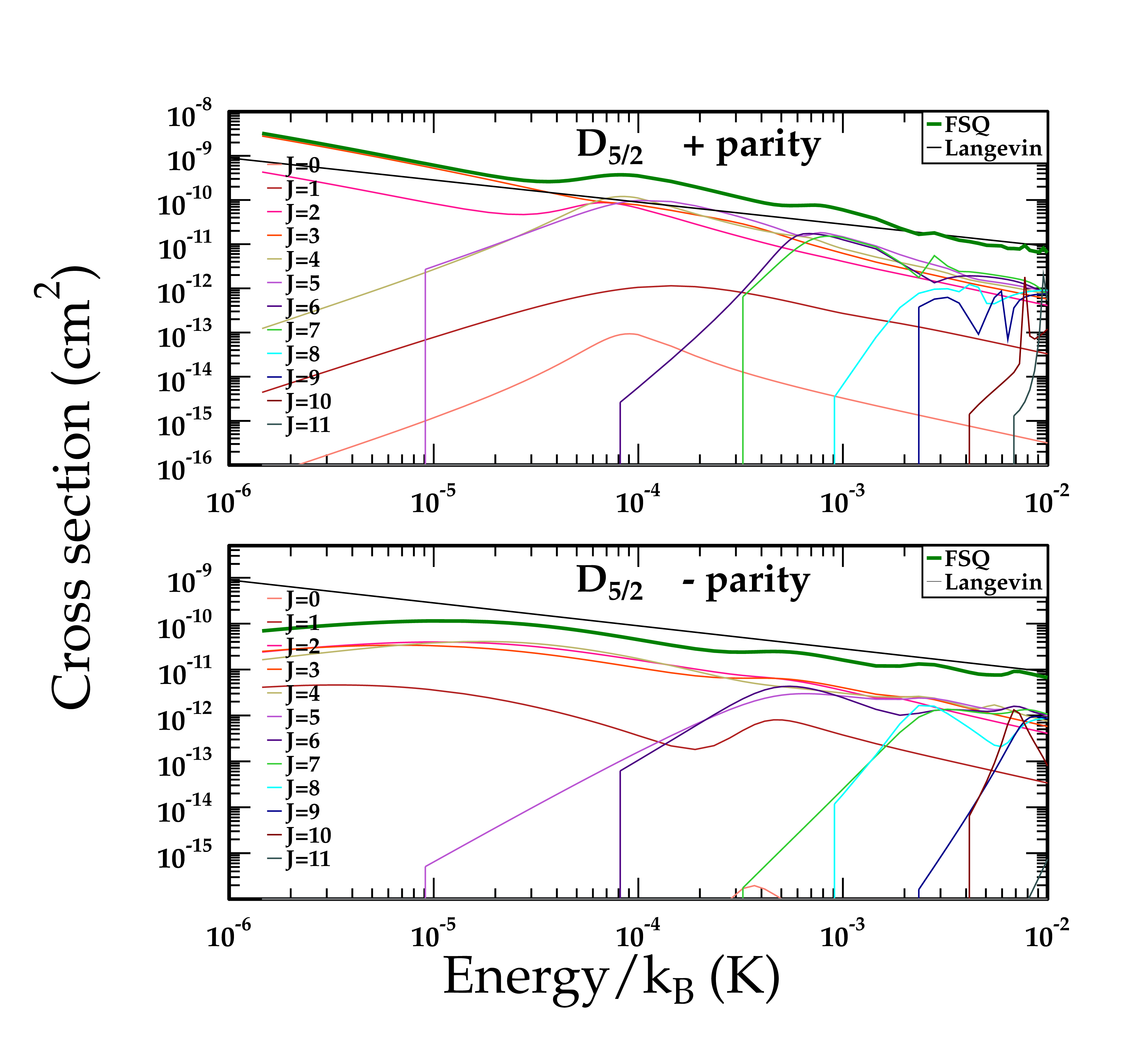}
 \caption{The parity dependent (thick green line) and the most relevant $J$-dependent partial cross sections (thin lines) for the FSQ process as a function of the temperature, for the $D_{5/2}$ incoming channels. For the + parity states at low temperatures $J=3$ defines the character of the cross section, while at higher temperatures $J\, =\,4$ then $J\,=\,6$ becomes dominant. For the - parity states at low temperatures $J=2$ and $J\,=\,3$ defines the character of the cross section, while at higher temperatures $J\,=\,6$ and $J\,=\, 8 $ becomes dominant.}
 \label{fig:xsectionJFSQ52}
\end{figure}

For $^2D_{5/2}$ incoming channels and $+$ parity (Fig. \ref{fig:xsectionJFSQ52}), we find a prominent shape resonance for both FSQ and NRCE processes around $10^{-4}$ K. In case of FSQ the resonance is mainly generated by $J\,=\, 4,\,5$. Regarding the NRCE, only even $J$ values contribute to the  process, and $J\,=\,2,\,4$ are responsible for the appearance of the resonance. Due to the outstanding shape resonance the cross section for the FSQ process is slightly larger than the Langevin one. In case of $-$ parity the FSQ process also dominates the NRCE process for which only odd $J$ values contribute, and the cross section for the NRQ process is smaller with more than two orders of magnitude. 

\begin{figure}[]
 \centering
 \includegraphics[width=8 cm]{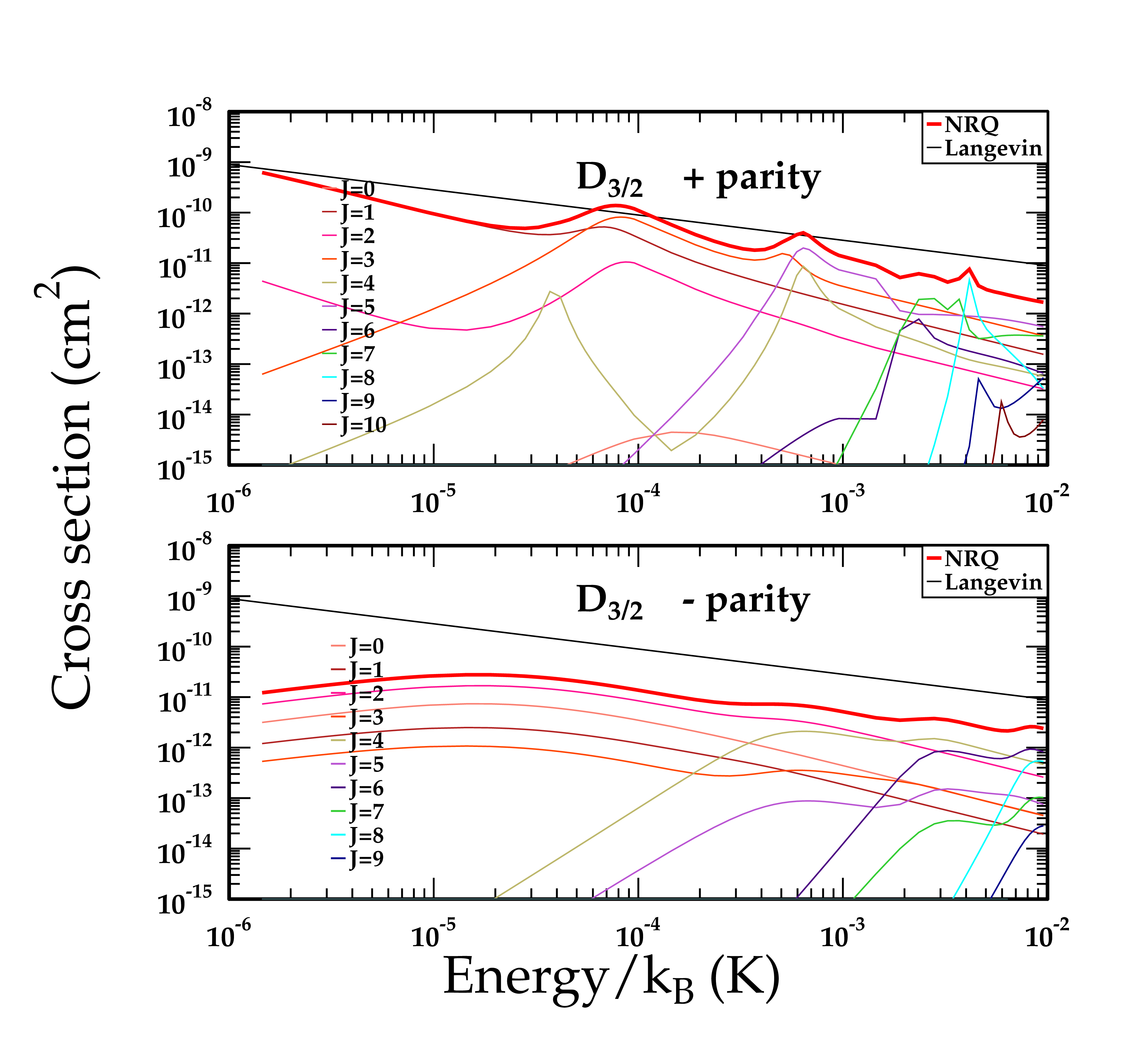}
 \caption{The parity dependent (thick red line) and the most relevant  $J$-dependent partial cross sections (thin lines) for the NRQ process as a function of the temperature, for the $D_{3/2}$ incoming channels. For the + parity states at low temperatures $J=1$ defines the character of the total cross section, while at higher temperatures $J\, =\,3$ then $J\,=\,5$ and $J\,=\,8$ becomes dominant. For the - parity states at low temperatures $J=2$ defines the character of the cross section, while at higher temperatures $J\,=\,4$ and $J\,=\, 6 $ becomes dominant. }
 \label{fig:xsectionJNRQ32}
\end{figure}

Regarding the $^2D_{3/2}$ incoming channels (Fig. \ref{fig:xsectionJNRQ32}), and for $+$ parity up to $10^{-5} K$ the cross sections have similar behaviour. In case of NRQ process the structure appearing around $10^{-4} K$ is created mainly by $J =1,3$, while around $10^{-3} K$ by $J=5$. For the NRCE process the structure appearing at $4\times10^{-5} K$ is the result of $J=4$. 

\begin{figure}[]
 \centering
 \includegraphics[width=8 cm]{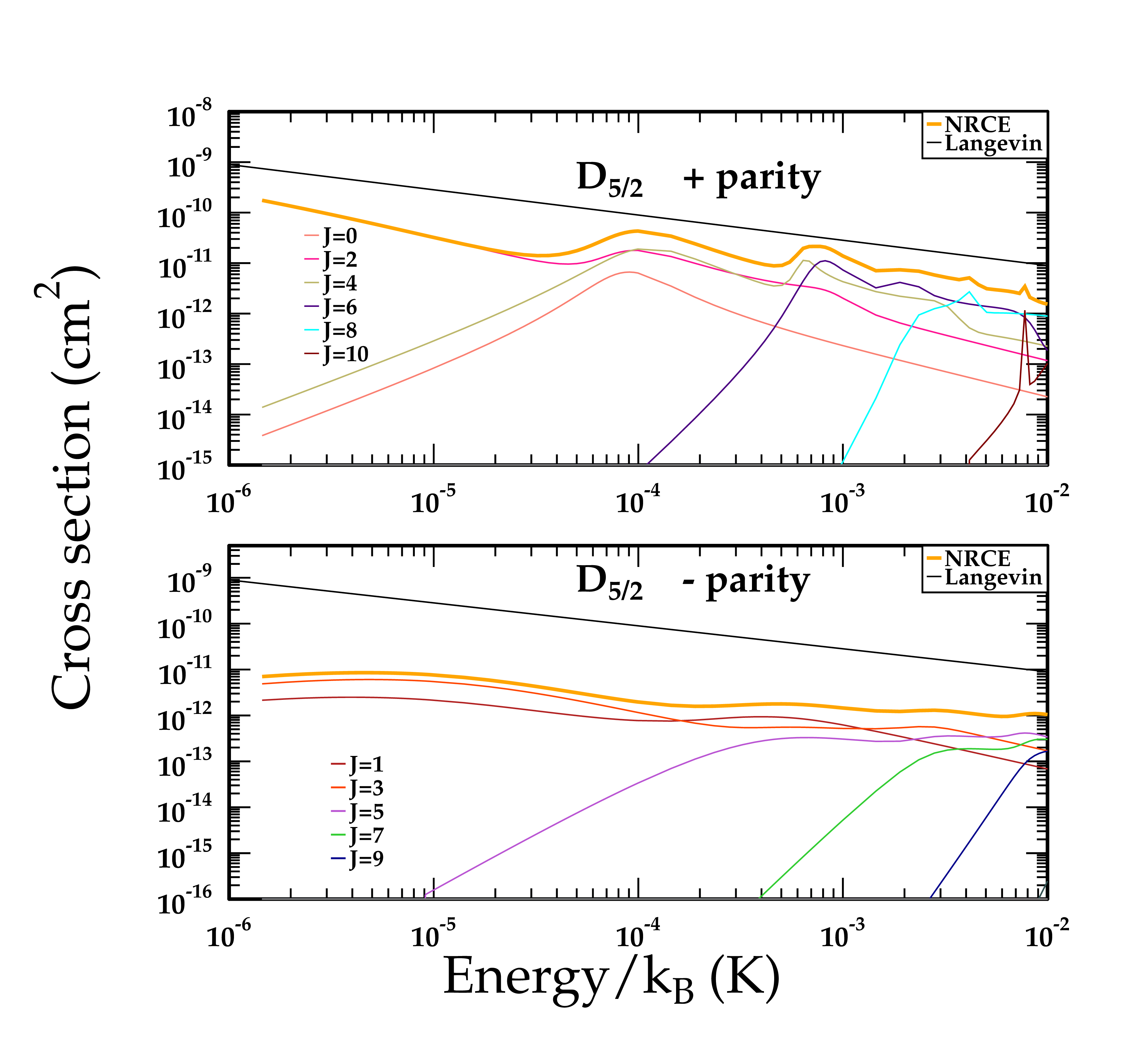}
 \caption{The parity dependent (thick yellow line) and the most relevant  $J$-dependent partial cross sections (thin lines) for the NRCE process as a function of the temperature, for the $D_{5/2}$ incoming channels. For the + parity states at low temperatures $J=2$ defines the character of the cross section, while at higher temperatures $J\, =\,4$ then $J\,=\,6$ becomes dominant. For the - parity states at low temperatures $J=3$ and $J\,=\,1$ defines the character of the cross section, while at higher temperatures $J\,=\,5$ becomes dominant. }
 \label{fig:xsectionJNRCE52}
\end{figure}

\begin{figure}[]
 \centering
 \includegraphics[width=8 cm]{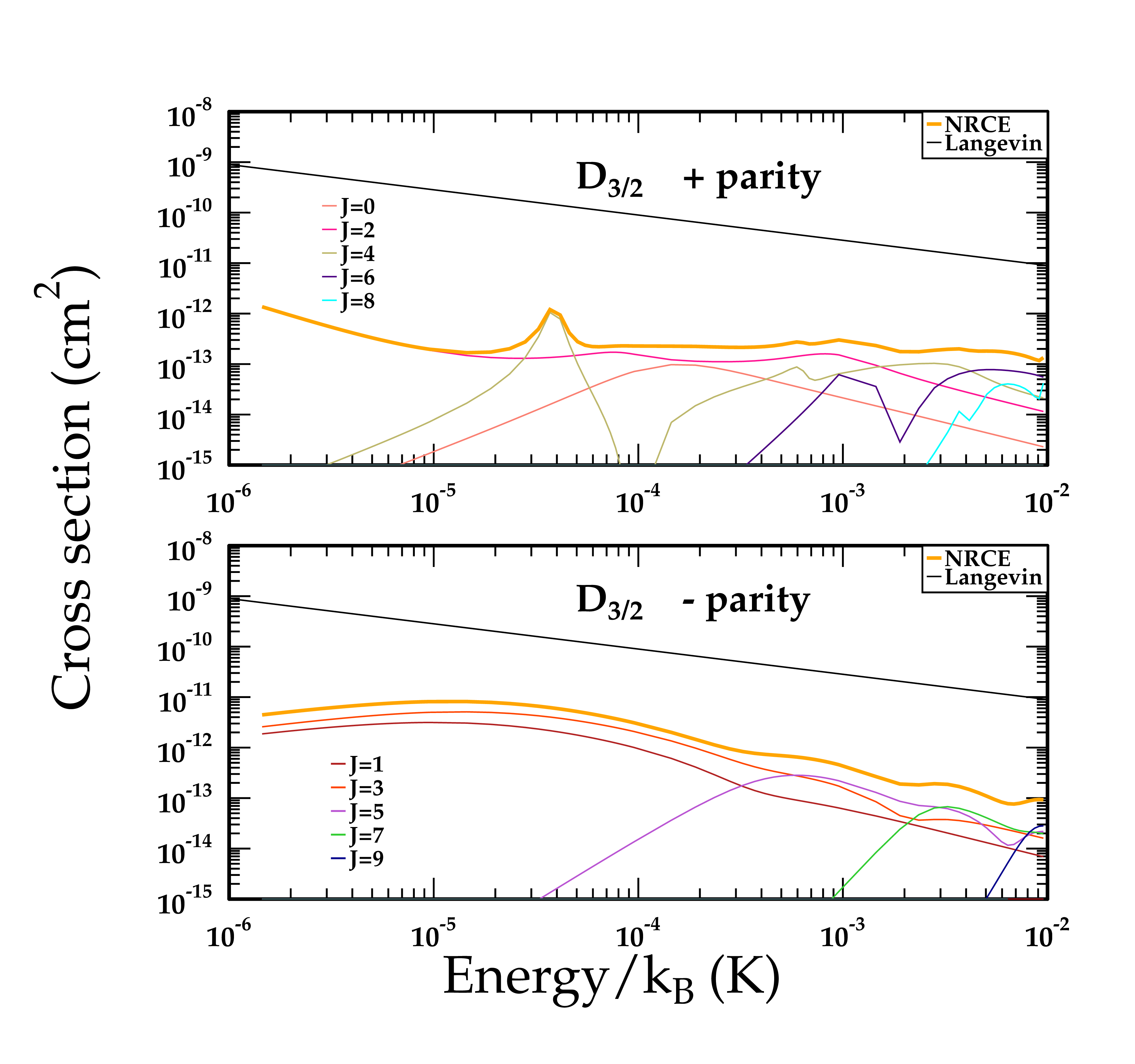}
 \caption{The parity dependent (thick yellow line) and the most relevant $J$-dependent partial cross sections (thin lines) for the NRCE process as a function of the temperature, for the $D_{3/2}$ incoming channels. For the + parity states almost through the whole temperature range $J\,=\,2$ and $J\,=\,4 $ defines the character of the cross section. For the - parity states $J\,=\,3$ then $J\,=5 $ and $J\,=\,7$ becomes dominant. We note here that the NRCE process  concerns only channels correlating to the ${\rm Li}^{+}\, + \, {\rm Ba}(^{1}S)$ asymptote, which has exclusively $^{1}\Sigma ^{+}$ character, thus for $+$ parity case only even $J$, while for $-$ parity case only odd $J$ values contribute to the total cross section.
 }
 \label{xsectionJNRCE32}
\end{figure}


\clearpage

\begin{thebibliography}{10}

\bibitem{baroni2024}
C.~Baroni, G.~Lamporesi, and M.~Zaccanti.
\newblock Quantum mixtures of ultracold atomic gases.
\newblock arxiv:2405.14562, 2024.

\bibitem{lous2022}
R.~S. Lous and R.~Gerristma.
\newblock Chapter two - ultracold ion-atom experiments: cooling, chemistry, and
  quantum effects.
\newblock {\em Adv. At. Mol. Phys.}, 71:65, 2022.

\bibitem{deiss2024}
M.~Dei{\ss}, S.~Willitsch, and J.~Hecker Denschlag.
\newblock Cold trapped molecular ions and hybrid platforms for ions and neutral
  particles.
\newblock {\em Nat. Phys.}, 20:713, 2024.

\bibitem{krukow2016a}
Artjom Kr\"ukow, Amir Mohammadi, Arne H\"arter, Johannes~Hecker Denschlag,
  Jes\'us P\'erez-R\'{\i}os, and Chris~H. Greene.
\newblock Energy scaling of cold atom-atom-ion three-body recombination.
\newblock {\em Phys. Rev. Lett.}, 116:193201, 2016.

\bibitem{dieterle2020}
Thomas Dieterle, Moritz Berngruber, Christian H{\"o}lzl, Robert L{\"o}w,
  Krzysztof Jachymski, Tilman Pfau, and Florian Meinert.
\newblock Inelastic collision dynamics of a single cold ion immersed in a
  bose-einstein condensate.
\newblock {\em Phys. Rev. A}, 102(4):041301, 2020.

\bibitem{mohammadi2021}
Amir Mohammadi, Artjom Kr\"ukow, Amir Mahdian, Markus Dei\ss{}, Jes\'us
  P\'erez-R\'{\i}os, Humberto da~Silva, Maurice Raoult, Olivier Dulieu, and
  Johannes Hecker~Denschlag.
\newblock Life and death of a cold {B}a{R}b$^{+}$ molecule inside an ultracold
  cloud of {R}b atoms.
\newblock {\em Phys. Rev. Research}, 3:013196, 2021.

\bibitem{perez-rios2021}
Jes{\'u}s P{\'e}rez-R{\'\i}os.
\newblock Cold chemistry: a few-body perspective on impurity physics of a
  single ion in an ultracold bath.
\newblock {\em Molec. Phys.}, 119(8):e1881637, 2021.

\bibitem{chowdhury2024}
Saajid Chowdhury and Jesús Perez‐Ríos.
\newblock Ion solvation in atomic baths: From snowballs to polarons.
\newblock {\em Natural Sciences}, May 2024.

\bibitem{hirzler2023}
H.~Hirzler, E.~Trimby, R.~Gerritsma, A.~Safavi-Naini, and J.~P\'erez-R\'ios.
\newblock Trap-assisted complexes in cold atom-ion collisions.
\newblock {\em Phys. Rev. Lett.}, 130:143003, 2023.

\bibitem{grier2009}
Andrew~T. Grier, Marko Cetina, Fedja Oru\ifmmode \check{c}\else
  \v{c}\fi{}evi\ifmmode~\acute{c}\else \'{c}\fi{}, and Vladan
  Vuleti\ifmmode~\acute{c}\else \'{c}\fi{}.
\newblock Observation of cold collisions between trapped ions and trapped
  atoms.
\newblock {\em Phys. Rev. Lett.}, 102:223201, 2009.

\bibitem{haerter2012}
A.~H\"arter, A.~Kr\"ukow, A.~Brunner, W.~Schnitzler, S.~Schmid, and J.~Hecker
  Denschlag.
\newblock Single ion as a three-body reaction center in an ultracold atomic
  gas.
\newblock {\em Phys. Rev. Lett.}, 109:123201, 2012.

\bibitem{dutta2018}
Sourav Dutta and S.~A. Rangwala.
\newblock Cooling of trapped ions by resonant charge exchange.
\newblock {\em Phys. Rev. A}, 97:041401, 2018.

\bibitem{smith2005}
W.~W. Smith, O.~Makarov, and J.~Lin.
\newblock Cold ion–neutral collisions in a hybrid trap.
\newblock {\em J. Mod. Opt.}, 52:2253, 2005.

\bibitem{schmid2010}
S.~Schmid, A.~H\"arter, and J.~Hecker Denschlag.
\newblock Dynamics of a cold trapped ion in a {B}ose-{E}instein condensate.
\newblock {\em Phys. Rev. Lett.}, 105:133202, 2010.

\bibitem{hall2011}
F.~H.~J. Hall, M.~Aymar, N.~Bouloufa, O.~Dulieu, and S.~Willitsch.
\newblock Light-assisted ion-neutral reactive processes in the cold regime :
  radiative molecule formation vs. charge exchange.
\newblock {\em Phys. Rev. Lett.}, 107:243202, 2011.

\bibitem{ratschbacher2012}
L.~Ratschbacher, C.~Zipkes, C.~Sias, and Michael K\"ohl.
\newblock Controlling chemical reactions of a single particle.
\newblock {\em Nature Phys.}, 8:649, 2012.

\bibitem{haze2013}
S.~Haze, S.~Hata, M.~Fujinaga, and T.~Mukaiyama.
\newblock Observation of elastic collisions between lithium atoms and calcium
  ions.
\newblock {\em Phys. Rev. A}, 87:052715, 2013.

\bibitem{hall2013b}
F.~H.J. Hall, M.~Aymar, M.~Raoult, O.~Dulieu, and S.~Willitsch.
\newblock Light-assisted cold chemical reactions of barium ions with rubidium
  atoms.
\newblock {\em Molec. Phys.}, 111:1683, 2013.

\bibitem{joger2017}
J.~Joger, H.~F\"urst, N.~Ewald, T.~Feldker, M.~Tomza, and R.~Gerritsma.
\newblock Observation of collisions between cold {L}i atoms and {Y}b$^{+}$
  ions.
\newblock {\em Phys. Rev. A}, 96:030703, 2017.

\bibitem{sikorsky2018}
Tomas Sikorsky, Ruti Ben-shlomi Ziv~Meir, Nitzan Akerman, and Roee Ozeri.
\newblock Spin-controlled atom–ion chemistry.
\newblock {\em Nat. Commun.}, 9:920, 2018.

\bibitem{mills2019}
Michael Mills, Prateek Puri, Ming Li, Steven~J. Schowalter, Alexander Dunning,
  Christian Schneider, Svetlana Kotochigova, and Eric~R. Hudson.
\newblock Engineering excited-state interactions at ultracold temperatures.
\newblock {\em Phys. Rev. Lett.}, 122:233401, 2019.

\bibitem{li2020}
H.~Li, S.~Jyothi, M.~Li, J.~Klos, A.~Petrov, Brown, and S.~K.R., Kotochigova.
\newblock Photon-mediated charge exchange reactions between $^{39}${K} atoms
  and $^{40}${C}a$^+$ ions in a hybrid trap.
\newblock {\em Phys. Chem. Chem. Phys.}, 22:10870, 2020.

\bibitem{schmidt2020}
J.~Schmidt, P.~Weckesser, F.~Thielemann, T.~Schaetz, and L.~Karpa.
\newblock Optical traps for sympathetic cooling of ions with ultracold neutral
  atoms.
\newblock {\em Phys. Rev. Lett.}, 124:053402, 2020.

\bibitem{hall2013a}
F.~H.J. Hall, P.~Eberle, G.~Hegi, M.~Raoult, M.~Aymar, O.~Dulieu, and
  S.~Willitsch.
\newblock Ion-neutral chemistry at ultralow energies: dynamics of reactive
  collisions between laser-cooled {C}a$^{+}$ ions and {R}b atoms in an ion-atom
  hybrid trap.
\newblock {\em Molec. Phys.}, 111:2020, 2013.

\bibitem{saito2017}
R.~Saito, S.~Haze, M.~Sasakawa, R.~Nakai, M.~Raoult, H.~Da~Silva, O.~Dulieu,
  and T.~Mukaiyama.
\newblock Characterization of charge-exchange collisions between ultracold
  $^{6}\mathrm{Li}$ atoms and ${^{40}\mathrm{Ca}}^{+}$ ions.
\newblock {\em Phys. Rev. A}, 95:032709, 2017.

\bibitem{benshlomi2020}
Ruti Ben-shlomi, Romain Vexiau, Ziv Meir, Tomas Sikorsky, Nitzan Akerman,
  Meirav Pinkas, Olivier Dulieu, and Roee Ozeri.
\newblock Direct observation of ultracold atom-ion excitation exchange.
\newblock {\em Phys. Rev. A}, 102:031301, 2020.

\bibitem{xing2022}
X.~Xing, H.~da~Silva~Jr., R.~Vexiau, N.~Bouloufa-Maafa, S.~Willitsch, and
  O.~Dulieu.
\newblock Ion-loss events in a hybrid trap of cold {R}b-{C}a$^+$:
  Photodissociation, blackbody radiation, and nonradiative charge exchange.
\newblock {\em Phys. Rev. A}, 106:062609, 2022.

\bibitem{cetina2012}
Marko Cetina, Andrew~T. Grier, and Vladan Vuleti\ifmmode~\acute{c}\else
  \'{c}\fi{}.
\newblock Micromotion-induced limit to atom-ion sympathetic cooling in paul
  traps.
\newblock {\em Phys. Rev. Lett.}, 109:253201, 2012.

\bibitem{feldker2020}
T.~Feldker, H.~F\"urst, H.~Hirzler, N.~V. Ewald, M.~Mazzanti, D.~Wiater,
  M.~Tomza, and R.~Gerritsma.
\newblock Buffer gas cooling of a trapped ion to the quantum regime.
\newblock {\em Nat. Phys.}, 16:413, 2020.

\bibitem{tomza2015}
Micha\l{} Tomza, Christiane~P. Koch, and Robert Moszynski.
\newblock Cold interactions between an {Y}b$^{+}$ ion and a {L}i atom:
  Prospects for sympathetic cooling, radiative association, and feshbach
  resonances.
\newblock {\em Phys. Rev. A}, 91:042706, 2015.

\bibitem{weckesser2021b}
P.~Weckesser, F.~Thielemann, D.~Wiater, A.~Wojciechowska, L.~Karpa,
  K.~Jachymski, M.~Tomza, T.~Walker, and T.~Schaetz.
\newblock Observation of {F}eshbach resonances between a single ion and
  ultracold atoms.
\newblock {\em Nature}, 600:429, 2021.

\bibitem{thielemann2024}
F.~Thielemann, J.~Siemund, D.~von Schoenfeld, W.~Wu, P.~Weckesser,
  K.~Jachymski, T.~Walker, and T.~Schaetz.
\newblock Exploring atom-ion {F}eshbach resonances below the s-wave limit.
\newblock arXiv:2406.13410, 2024.

\bibitem{hirzler2022}
H.~Hirzler, R.~S. Lous, E.~Trimby, J.~P\'erez-R\'ios, A.~Safavi-Naini, and
  R.~Gerritsma.
\newblock Observation of chemical reactions between a trapped ion and ultracold
  {F}eshbach dimers.
\newblock {\em Phys. Rev. Lett.}, 128:103401, 2022.

\bibitem{haze2015}
Shinsuke Haze, Ryoichi Saito, Munekazu Fujinaga, and Takashi Mukaiyama.
\newblock Charge-exchange collisions between ultracold fermionic lithium atoms
  and calcium ions.
\newblock {\em Phys. Rev. A}, 91:032709, 2015.

\bibitem{weckesser2021a}
P.~Weckesser, F.~Thielemann, D.~Hoenig, A.~Lambrecht, L.~Karpa, and T.~Schaetz.
\newblock Trapping, shaping, and isolating of an ion {C}oulomb crystal via
  state-selective optical potentials.
\newblock {\em Phys. Rev. A}, 103(1):013112, 2021.

\bibitem{leschhorn2012}
G.~Leschhorn, T.~Hasegawa, and T.~Schaetz.
\newblock Efficient photo-ionization for barium ion trapping using a
  dipole-allowed resonant two-photon transition.
\newblock {\em Appl. Phys. B}, 108:159, 2012.

\bibitem{aymar2005}
M.~Aymar and O.~Dulieu.
\newblock Calculation of accurate permanent dipole moments of the lowest
  $^{1,3}{\Sigma}^+$ states of heteronuclear alkali dimers using extended basis
  sets.
\newblock {\em J. Chem. Phys.}, 122:204302, 2005.

\bibitem{aymar2011}
M.~Aymar, R.~Gu\'erout, and O.~Dulieu.
\newblock Structure of the alkali-metal-atom-strontium molecular ions : towards
  photoassociation and formation of cold molecular ions.
\newblock {\em J. Chem. Phys.}, 135:064305, 2011.

\bibitem{aymar2012}
M.~Aymar and O.~Dulieu.
\newblock The electronic structure of the alkaline-earth-atom (ca, sr, ba)
  hydride molecular ions.
\newblock {\em J. Phys. B}, 45:215103, 2012.

\bibitem{deiglmayr2008}
J.~Deiglmayr, M.~Aymar, R.~Wester, M.~Weidem\"uller, and O.~Dulieu.
\newblock Calculations of static dipole polarizabilities of alkali dimers:
  Prospects for alignment of ultracold molecules.
\newblock {\em J. Chem. Phys.}, 129:064309, 2008.

\bibitem{sardar2023}
Dibyendu Sardar and Somnath Naskar.
\newblock Cold collisions between alkali metals and alkaline-earth metals in
  the heteronuclear atom-ion system $\mathrm{Li}+{B}a^{+}$.
\newblock {\em Phys. Rev. A}, 107:043323, 2023.

\bibitem{akkari2024}
S.~Akkari, W.~Zrafi, H.~Ladjimi, M.~Bejaoui, J.~Dhiflaoui, and H.~Berriche.
\newblock Electronic structure of ground and low-lying excited states of
  {B}a{L}i$^{+}$ molecular ion: spin-orbit effect, radiative lifetimes and
  {F}ranck-{C}ondon factor.
\newblock {\em Phys. Scripta}, 99:035403, 2024.

\bibitem{cimiraglia1985}
R~Cimiraglia, J~P Malrieu, M~Persico, and F~Spiegelmann.
\newblock Quasi-diabatic states and dynamical couplings from ab initio {CI}
  calculations: a new proposal.
\newblock {\em J. Phys. B}, 18:3073, 1985.

\bibitem{angeli1996}
C.~Angeli and M.~Persico.
\newblock Quasi-diabatic and adiabatic states and potential energy curves for
  {N}a-{C}d collisions and excimer formation.
\newblock {\em Chem. Phys.}, 204:57, 1996.

\bibitem{landau1967}
L.~Landau and E.~Lifchitz.
\newblock {\em M\'{e}canique Quantique - {T}h\'{e}orie non-relativiste}.
\newblock Editions Mir, 1967.

\bibitem{gao2013}
Bo~Gao.
\newblock Quantum-defect theory for $-1/{r}^{4}$-type interactions.
\newblock {\em Phys. Rev. A}, 88:022701, 2013.

\bibitem{langevin1905}
P.~Langevin.
\newblock {\em Ann. Chim. Phys.}, 5:245, 1905.

\bibitem{tscherbul2016}
T.~V. Tscherbul, P.~Brumer, and A.~A. Buchachenko.
\newblock Spin-orbit interactions and quantum spin dynamics in cold ion-atom
  collisions.
\newblock {\em Phys. Rev. Lett.}, 117:143201, 2016.

\bibitem{dion2001}
C.~M. Dion, C.~Drag, O.~Dulieu, B.~Laburthe Tolra, F.~Masnou-Seeuws, and
  P.~Pillet.
\newblock Resonant coupling in the formation of ultracold ground state
  molecules via photoassociation.
\newblock {\em Phys. Rev. Lett.}, 86:2253, 2001.

\bibitem{LiBa-ion}
Pascal Weckesser, Fabian Thielemann, Daniel Hoenig, Alexander Lambrecht, Leon
  Karpa, and Tobias Schaetz.
\newblock Trapping, shaping, and isolating of an ion coulomb crystal via
  state-selective optical potentials.
\newblock {\em Physical Review A}, 103(1):013112, 2021.

\bibitem{berkeland1998minimization}
DJ~Berkeland, JD~Miller, James~C Bergquist, Wayne~M Itano, and David~J
  Wineland.
\newblock Minimization of ion micromotion in a {P}aul trap.
\newblock {\em Journal of applied physics}, 83(10):5025--5033, 1998.

\bibitem{johnson1973}
B.R Johnson.
\newblock The multichannel log-derivative method for scattering calculations.
\newblock {\em J. Comp. Phys.}, 13:445, 1973.

\bibitem{alexander1987}
M.~H. Alexander and D.~E. Manolopoulos.
\newblock A stable linear reference potential algorithm for solution of the
  quantum close-coupled equations in molecular scattering theory.
\newblock {\em J. Chem. Phys.}, 86:2044, 1987.

\end{thebibliography}

\end{document}